\documentclass{jpp}
\usepackage{graphicx}
\usepackage[T1]{fontenc}
\usepackage{amsmath}
\newcommand{\partder}[2]{\frac{\partial #1}{\partial #2}}
\newcommand{\der}[2]{\frac{d #1}{d #2}}
\usepackage{bm}
\usepackage{cancel}
\usepackage{xcolor}
\usepackage{subcaption}

 \definecolor{red}{rgb}{1.0,0.0,0.0}
 \definecolor{gre}{rgb}{0.0,1.0,0.0}
 \definecolor{blu}{rgb}{0.0,0.0,1.0}

\usepackage{epstopdf}
\AppendGraphicsExtensions{.tiff}
 
\title{Heat conduction in an irregular magnetic ﬁeld: Part II. Heat transport as a measure of the effective non-integrable volume}
\shorttitle{Heat transport as a measure of the effective non-integrable volume}

\author{Elizabeth J. Paul\aff{1,2}\corresp{\email{epaul@princeton.edu}}, Stuart R. Hudson\aff{2}, Per Helander\aff{3}}
\affiliation{\aff{1}Department of Astrophysical Sciences, Princeton University,
Princeton, NJ 08544, USA
\aff{2}Princeton Plasma Physics Laboratory, PO Box 451, Princeton NJ 08543, USA
\aff{3}Max Planck Institute for Plasma Physics, Greifswald, Germany} 

\begin{document}

\maketitle

\begin{abstract}
Given the large anisotropy of transport processes in magnetized plasmas, the magnetic field structure can strongly impact heat diffusion: magnetic surfaces and cantori form barriers to transport while chaotic layers and island structures can degrade confinement. When a small but {\color{black}non-zero} amount of perpendicular diffusion is included, the structure of the magnetic field becomes less important, allowing pressure gradients to be supported across chaotic regions and island chains. We introduce a metric for the effective volume {\color{black}over which the local parallel diffusion dominates} based on the solution to the anisotropic heat diffusion equation. To validate this metric, we consider model fields with a single island chain and a strongly chaotic layer for which analytic predictions of the relative parallel and perpendicular transport can be made. We also analyze critically chaotic fields produced from different sets of perturbations, highlighting the impact of the mode number spectrum on the heat transport. {\color{black} Our results indicate that this metric coincides with the effective volume of non-integrability in the limit $\kappa_{\perp} \rightarrow 0$.} We propose that this metric be used to assess the impact of non-integrability on the heat transport in stellarator equilibria.
\end{abstract}

\section{Introduction}

The heat transport in irregular magnetic fields is critical for understanding magnetic confinement experiments and astrophysical systems. Tokamak experiments exploit control coils which produce resonant magnetic perturbations for control of edge localized modes (ELMs) \citep{Evans2005}. Such perturbations must induce a small amount of stochasticization near an edge resonant surface to suppress ELMs without significant degradation of the core confinement. These coils can impart structure in the field outside the last-closed flux surface, resulting in homoclinic tangles \citep{Evans2005} and stochastic layers \citep{Punjabi2008} which can impact the heat flux deposition at the divertor. Reversed Field Pinches (RFPs) provides confinement through magnetic fields generated largely by the plasma current. The transport in an RFP is often dominated by parallel diffusion due to stochastic magnetic fields which form due to tearing mode activity \citep{Sarff2013}. The presence of a transport barrier \citep{Biewer2003,Lorenzini2009} can enable the support of a temperature gradient. 
As stellarators provide confinement with three-dimensional magnetic fields, the impact of non-integrability is critical for heat transport in both the core and divertor regions. For example, the stochastic edge of LHD features island remnants, stochastic layers, and partial barriers. Due to the long connection lengths, both the parallel and perpendicular transport are important for determining the heat flux impacting the divertor plates in a stellarator \citep{Feng2011}. Stochastic magnetic fields are also expected to exist in several astrophysical environments. For example, the stochasticity of the galactic magnetic field may impact the transport of cosmic rays \citep{Jokipii1969,Chuvilgin1993}.

Given the Hamiltonian nature of magnetic field line flow \citep{Cary1983}, there are a number of tools from the field of dynamical systems that enable the quantification of the transition to chaos. At the most granular level, the linear stability of a single resonance is quantified through Greene's residue \citep{Greene1979}, $R$, where $R < 0$ or $R>1$ indicates unstable (hyperbolic or hyperbolic-with-reflection) orbits and $R \in [0,1]$ indicates stable (elliptic) orbits. 
Thus one technique to reduce the stochasticity and island size in a stellarator is through minimization of the magnitude of the residue \citep{Hanson1984,Cary1986}. Similarly, the island width under the small island approximation can be evaluated through a linearization about the island center \citep{Cary1991}. A shortcoming of these approaches is their limitation to a single resonance rather than a volumetric measure of non-integrability. 

One can also evaluate the existence or non-existence of a KAM surface to determine the transition to chaos in a given region. To estimate the value of the perturbation required to destroy the last KAM surface between two primary resonances, one may compute the Chirikov overlap criterion \citep{Chirikov1979}, obtained from the overlap of resonances and their second order island chains.
Greene has also conjectured a procedure for determining the existence of a KAM surface by evaluating the residues of the high-order rational convergents of a given irrational. By examining the stability of the neighboring periodic orbits to the most stable KAM surface, one can determine the transition to stochasticity \citep{Greene1979,Mackay1992,Falcolini1992}. In a similar way, one can evaluate Mather's $\Delta W$, the difference in action between the X and O points associated with a given rational rotation number, $p/q$. If $\Delta W \rightarrow 0$ as $p/q$ approaches a given irrational, then the corresponding KAM surface survives \citep{Mather1986}. 
Furthermore, converse KAM theory \citep{Mather1984,Mackay1985,Mackay2018} provides a sufficient condition for non-existence of a KAM surface by considering the rotation of nearby trajectories with respect to a given foliation. 

While each of these techniques may shed light on the existence or non-existence of a KAM surface in a given region, another set of techniques are required to provide insight into the relative impact of non-integrability on the transport in a dynamical system \citep{Meiss2015}. For example, Mather's $\Delta W$ is related to the flux of trajectories across a given curve in phase space, providing a bound on the transport \citep{Mackay1984}. One can then partition phase space into subvolumes bounded by KAM surfaces or cantori and model transport between states, such as with a Markov model \citep{Meiss1986}.


Rather than using nonlinear dynamical surrogates of transport, one could directly evaluate the transport in a given magnetic field line flow. A logical tactic for studying the transport is by introducing a tracer, such as the temperature, $T(\textbf{x})$, which diffuses along the magnetic field subject to some initial condition $T_0(\textbf{x})$,
{\color{black}
\begin{subequations}
\begin{align}
    \partder{T(\textbf{x},t)}{t} - \nabla \cdot \left( \kappa_{\|} \hat{\textbf{b}} \hat{\textbf{b}} \cdot \nabla T \right) &= 0 \\
    T(\textbf{x},0) &= T_0(\textbf{x}).
\end{align}
\label{eq:parallel_transport}
\end{subequations}
}
The temperature at a given point, $T(\textbf{x},t)$, can be computed by integrating over all contributions of the initial condition along the magnetic field line path \citep{Del2012}. In the steady-state $t \rightarrow \infty$ limit, the temperature becomes constant on magnetic field lines; it is simply the average of the initial condition along the magnetic field line path,
{\color{black}
\begin{align}
    T(\textbf{x},t \rightarrow \infty) = \lim_{L \rightarrow \infty} \frac{\int_{0}^{L} dl \, T_0(\textbf{x}(l))}{\int_{0}^{L} dl },
\end{align}
}
where $l$ measures length along a field line. Because the temperature is constant along field lines, in a chaotic magnetic field this leads to a fractal-like structure in the temperature profile \citep{Hudson2008,Del2012}. In this way, local flattening of the temperature profile can be used as a measure of the local stochasticity.



While the structure of the magnetic field itself can give rise to transport, in a plasma a {\color{black}non-zero} amount of perpendicular transport exists due to the gyromotion, cross-field drifts, collisions, and fluctuations. Although the effective perpendicular diffusion can be extremely small in a magnetic confinement device (experimental measurements on the TEXTOR, DIII-D, and ASDEX tokamaks indicate $\kappa_{\perp}/\kappa_{\|} \sim 10^{-9}-10^{-7}$ \citep{Meskat2001,Holzl2009,Bardoczi2016}, where $\kappa_{\perp}$ is the perpendicular diffusion coefficient and $\kappa_{\|}$ is the parallel diffusion coefficient), {\color{black}non-zero} perpendicular diffusion can substantially alter the heat transport. For example, in the Large Helical Device (LHD) a $\beta = 4.8\%$ discharge demonstrated a 32\% reduction in nested magnetic surfaces in comparison with the vacuum magnetic field. However, significant flattening of the temperature profile was not observed \citep{Sakakibara2008}, indicating the relative importance of the perpendicular transport. The perpendicular diffusion becomes even more critical in the stochastic edge, where the collisionality is enhanced. 

\textcolor{black}{In the steady-state limit, the transport equation \eqref{eq:parallel_transport} can be modified to include the effect of perpendicular diffusion,
\begin{align}
    \nabla \cdot \left( \kappa_{\|}\nabla_{\|} T + \kappa_{\perp} \nabla_{\perp} T \right) &= 0, 
\end{align}
where the parallel gradient is $\nabla_{\|} =\hat{\textbf{b}} \hat{\textbf{b}} \cdot \nabla$ and the perpendicular gradient is $\nabla_{\perp} = \nabla - \nabla_{\|}$.} We will discuss important properties of this anisotropic diffusion equation (ADE) in \S \ref{sec:ade}. The relative importance of the structure of the magnetic field on transport can then be quantified by comparing the parallel diffusion,
{\color{black}
\begin{align}
    D_{\|} = \frac{1}{2} \int_{V} d^3 x \, \kappa_{\|} \left| \nabla_{\|} T\right|^2,
\end{align}
}
with the perpendicular diffusion,
{\color{black}
\begin{align}
    D_{\perp} = \frac{1}{2} \int_{V} d^3 x \, \kappa_{\perp} \left|\nabla_{\perp} T\right|^2,
\end{align}
}
where $V$ is the volume of the domain of interest. In this way, the non-integrability of the magnetic field can be quantified by its impact on the transport in a volume rather than through the analysis of individual resonances or the existence of a given KAM surface. This allows us to define an {\color{black}\textit{effective volume of perpendicular diffusion}} in \S \ref{sec:effective_volume}, {\color{black} which we conjecture converges to the \textit{effective volume of non-integrability} in the limit $\kappa_{\perp} \rightarrow 0$}. In Section \ref{sec:numerical} we present the numerical methods used to solve the diffusion equation and the model field used for our calculations. In Sections \ref{sec:single_island} and \ref{sec:strong_chaos} we perform calculations of the transport in the presence of a single island chain and a strongly chaotic layer to justify our metric. We then apply the metric to a critical chaotic layer created from resonances of different mode numbers in \S \ref{sec:critical_chaos}. We then conclude in \S \ref{sec:conclusions}.

\section{Properties of the anisotropic diffusion equation}
\label{sec:ade}

The anisotropic diffusion equation (ADE) can be expressed as,
\begin{align}
    \nabla \cdot \left(\bm{\kappa} \cdot \nabla T \right) &= 0,
\label{eq:anisotropic_diffusion}
\end{align}
with the diffusion tensor,
\begin{align}
    \bm{\kappa} = \kappa_{\perp}\textbf{I} + (\kappa_{\|} - \kappa_{\perp}) \hat{\textbf{b}} \hat{\textbf{b}},
    \label{eq:diffusion_tensor}
\end{align}
where the perpendicular, $\kappa_{\perp}$, and parallel, $\kappa_{\|}$, diffusion coefficients are taken to be constants. We compute solutions to \eqref{eq:anisotropic_diffusion} in an annular volume $\Omega$ bounded by two toroidal surfaces, $S_{-}$ and $S_{+}$ with constant Dirichlet boundary conditions $T_{-}$ and $T_{+}$. 

\subsection{Ellipticity}
We note that $\bm{\kappa}$ is positive definite for $\kappa_{\perp} > 0$ and $\kappa_{\|} \ge \kappa_{\perp}$ since,
\begin{align}
    \nabla T \cdot \bm{\kappa} \cdot \nabla T = \kappa_{\perp} |\nabla T|^2 + (\kappa_{\|}- \kappa_{\perp}) \left(\hat{\textbf{b}} \cdot \nabla T \right)^2 > 0,
    \label{eq:positive_definite}
\end{align}
for all $\nabla T \ne 0$. {\color{black} (While the assumption $\kappa_{\|} \ge \kappa_{\perp}$ is not necessary for ellipticity, it is sufficient according to \eqref{eq:positive_definite} and not too restrictive since we are interested in strong anisotropy, $\kappa_{\|}\gg \kappa_{\perp}$.)}
We can furthermore see that the anisotropic diffusion operator is elliptic, as it can be expressed as,
\begin{align}
    \sum_{i,j} \partial_i \left(\kappa_{ij} \partial_j T\right) = \sum_{i,j} \kappa_{ij} \partial_i  \partial_j T + \left(\partial_i \kappa_{ij}\right) \partial_j T.
    \label{eq:elliptic_problem}
\end{align}
Ellipticity of the anisotropic diffusion operator \eqref{eq:anisotropic_diffusion} follows from the symmetric positive-definiteness of the tensor which multiplies the second derivatives, $\kappa_{ij}$. Ellipticity implies the existence of a maximum principle (Appendix \ref{app:maximum_principle}),
\begin{align}
    \max_{\overline{\Omega}} T = \max_{\partial \Omega} T.
\end{align}
The corresponding statement also holds for the minimum. In words, the maximum (minimum) temperature must be obtained on the boundary. Furthermore, no local minima or maxima can occur within $\Omega$. {\color{black}As a further consequence of ellipticity, the topology of the isotherms is constrained (Appendix \ref{sec:topology}). Specifically, an isotherm cannot enclose a net volume, but instead must link the domain toroidal and poloidally as the boundaries (with the possible addition of a handle).}

\subsection{Variational principle}

The anisotropic diffusion equation can be obtained \citep{Helander2021}
from a variational principle involving the functional,
\begin{align}
    \mathcal{W}[T] = \frac{1}{2}\int_{\Omega} d^3 x \, \nabla T \cdot \bm{\kappa} \cdot \nabla T.
\end{align}
The first variation of $\mathcal{W}$ with respect to $T$ is computed to be,
\begin{align}
    \delta \mathcal{W}[\delta T] &= \int_{\Omega} d^3 x \, \nabla \delta T \cdot \bm{\kappa} \cdot \nabla T= - \int_{\Omega} d^3 x \, \delta T \nabla \cdot \left( \bm{\kappa} \cdot \nabla T \right),
\end{align}
upon application of the boundary condition $\delta T \rvert_{\partial \Omega} = 0$. Thus stationary points of $\mathcal{W}$ with respect to $T$ correspond to solutions of \eqref{eq:anisotropic_diffusion}. We can now compute the second variations as,
\begin{align}
    \delta^2\mathcal{W}[\delta T,\delta T'] = \int_{\Omega} d^3 x \, \nabla \delta T \cdot \bm{\kappa} \cdot \nabla \delta T'. 
\end{align}
We note that $\delta^2 \mathcal{W}[\delta T,\delta T] \ge 0$ under the assumption that $\kappa_{\perp}<\kappa_{\|}$. Therefore, $\mathcal{W}$ is a convex function, implying that any stationary point of $\mathcal{W}$ is a global minimum.
We will refer to the functional $\mathcal{W}$ that appears in this variational principle as the \textit{diffusion integral},
\begin{align}
    D = \frac{1}{2} \int_{\Omega} d^3 x \, \left( \kappa_{\|}| \nabla_\| T |^2 + \kappa_\perp |\nabla_\perp T |^2 \right),
\end{align}
as it is related to the entropy production functional \citep{Hameiri1987}. Thus the temperature profile {\color{black}minimizes} the entropy production subject to the boundary conditions.

We note that the total diffusion is related to the heat flux through the boundaries,
{\color{black}
\begin{align}
    Q = -\int_{S_+} d^2 x \, \hat{\textbf{n}} \cdot \textbf{q} = -\int_{S_-} d^2 x \, \hat{\textbf{n}} \cdot\textbf{q},
\end{align}
}
through
\begin{align}
    Q = \frac{2 D[T]}{T_{+} - T_{-}},
\end{align}
where $\textbf{q} = -\kappa_{\|} \nabla_{\|}T - \kappa_{\perp} \nabla_{\perp} T$ \textcolor{black}{and $\hat{\textbf{n}}$ is the unit normal oriented in the direction from $S_-$ to $S_+$}. This relation follows from,
\begin{align}
0 &= \int_\Omega d^3 x \, T \nabla \cdot {\bf q} = - \int_\Omega d^3 x \, {\bf q}  \cdot \nabla T  + \int_{S_+}  d^2 x \, T \hat{\textbf{n}} \cdot {\bf q} - \int_{S_-}  d^2 x \, T \hat{\textbf{n}} \cdot {\bf q} \\
&= 2 D[T] - Q (T_+ - T_-). \nonumber
\end{align}
Thus the total diffusion quantifies the heat flux required to support a given temperature gradient across a volume. In this work we will consider $T_+$ and $T_-$ to be fixed such that $D$ measures the required heat flux to support the prescribed temperature gradient. Since the perpendicular transport is relatively insensitive to the structure of the field, we can assess the impact of non-integrability on the transport through the local parallel diffusion, 
\begin{align}
    d_{\|} = \frac{1}{2}\kappa_{\|} |\nabla_{\|}T|^2,
    \label{eq:parallel_diffusion}
\end{align}
in comparison with the local perpendicular diffusion,
\begin{align}
     d_{\perp} = \frac{1}{2}\kappa_{\perp} |\nabla_{\perp}T|^2.
     \label{eq:perpendicular_diffusion}
\end{align}

\section{The effective volume of {\color{black}parallel diffusion}}
\label{sec:effective_volume}

Motivated by comparing $d_{\|}$ to $d_{\perp}$ in order to determine the impact of non-integrability on transport, we propose the following metric:
{\color{black}
\begin{align}
    \mathcal{V}_{\text{PD}} = \frac{\int_{\Omega} d^3 x \, \Theta \left(\kappa_{\|}|\nabla_{\|}T|^2-\kappa_{\perp} |\nabla_{\perp}T|^2\right)}{\int_{\Omega} d^3 x},
    \label{eq:effective volume}
\end{align}
}
where $\Theta$ is the Heaviside step function. In words, {\color{black}$\mathcal{V}_{\text{PD}}$} measures the fraction of volume over which the local parallel transport is larger than the perpendicular transport. While one could imagine many ways in which to compare the parallel and perpendicular diffusion, the above definition is easy to interpret as it is a positive scalar in $[0,1]$. 

{\color{black}While the definition of $\mathcal{V}_{\text{PD}}$ is not intrinsic to the field structure as it depends on $\kappa_{\perp}/\kappa_{\|}$, we can infer general trends in the limit of small $\kappa_{\perp}/\kappa_{\|}$.} In the limit of an integrable magnetic field with sufficiently small $\kappa_{\perp}$, the isotherms will {\color{black}largely} coincide with magnetic surfaces, $|\hat{\textbf{b}} \cdot \nabla T|$ will be relatively small, and thus the local parallel transport {\color{black}{will be}} negligible. 
 Thus {\color{black}$\mathcal{V}_{\text{PD}}$} will be {\color{black}small} in the limit of integrability. Since \textcolor{black}{the topology of the isotherms is constrained}, isotherms will not be able to completely align to the structure of the field in regions of non-integrability, and $|\hat{\textbf{b}} \cdot \nabla T|$ will be consequently larger, as measured by {\color{black}$\mathcal{V}_{\text{PD}}$}.

We can consider the metric {\color{black}$\mathcal{V}_{\text{PD}}$} to be a ``diffusive'' analog of a metric based on converse KAM theory. With a converse KAM approach, the non-existence of an invariant torus at a given point in phase space is determined from the rotation of nearby trajectories with respect to a chosen foliation. This allows one to determine regions of phase space over which invariant surfaces do not exist with the topology of the chosen foliation{\color{black}, i.e. a given class of tori} \citep{Mackay1985,Mackay2018}. Similarly, by comparing the local parallel and perpendicular diffusion, we can assess whether a given region of physical space is {\color{black}``effectively non-integrable'' (indicated by enhanced parallel diffusion)} with respect to its impact on the transport. This assessment similarly depends on the topology of the isotherms, which foliate the volume.
{\color{black} Since the isotherms must link the domain as the boundaries do, the ADE similarly provides an assessment of the non-integrability with respect to a limited class of tori.} For instance, an island chain will impact the parallel transport if the boundary surfaces are chosen to have the same topology as the unperturbed Hamiltonian, {\color{black} given that isotherms cannot have the same linking as the island}. If instead the boundary surfaces {\color{black}have the same linking} as the surfaces within the island chain, the parallel transport will not be impacted. 

To assess this metric, we consider two cases for which analytic predictions for the parallel transport can be employed: diffusion across a single island chain (\S \ref{sec:single_island}) and across a strongly chaotic layer (\S \ref{sec:strong_chaos}). In \S \ref{sec:critical_chaos} we then consider a model field for which such prediction is more challenging, namely a critically chaotic field with remnants of island chains and partial barriers.



\section{Numerical methods}
\label{sec:numerical}

\subsection{Linear solution}

We compute the numerical solution to the anisotropic diffusion equation \eqref{eq:anisotropic_diffusion}-\eqref{eq:diffusion_tensor} using a Fourier Galerkin discretization in $\theta$ and $\zeta$ and a fourth-order finite-difference radial discretization. The resulting sparse linear system is solved with the PETSc library \citep{petsc-web-page,petsc-user-ref,petsc-efficient} using the MUMPS package \citep{MUMPS:1,MUMPS:2} to compute the $LU$-factorization. Throughout we will normalize calculations such that $T_- = 0$, $T_+ = 1$, and $\kappa_{\|} = 1$.

\subsection{Model magnetic field}

We consider a model Hamiltonian,
{\color{black}
\begin{align}
    \chi(\psi,\theta,\zeta) = \frac{\iota' \psi^2}{2} + \sum_{m,n}\epsilon_{m,n} \psi (\psi - \overline{\psi}) \cos(m \theta - n \zeta),
\end{align}
}
which yields the magnetic field,
\begin{align}
    \textbf{B} = \nabla \psi \times \nabla \theta - \nabla \chi \times \nabla \zeta 
 \label{eq:model_field}  
\end{align}
where \textcolor{black}{$\psi = \overline{\psi} \rho/\overline{\rho}$} and $(\rho,\theta,\zeta)$ forms \textcolor{black}{an orthogonal} coordinate system \textcolor{black}{($\hat{\bm{\rho}} \times \hat{\bm{\theta}} \cdot \hat{\bm{\zeta}} = 1$)} with \textcolor{black}{$\rho \in [0,\overline{\rho}]$}, $\theta \in [0,2\pi)$, $\zeta \in [0,2\pi)$, \textcolor{black}{$|\nabla \rho| = 1$, $|\nabla \theta| = 1/\overline{\rho}$, and $|\nabla \zeta| = 1/L_{\zeta}$}. We will use the notation {\color{black}$\widetilde{\epsilon}_{m,n}(\psi) = \epsilon_{m,n}\psi(\psi-\overline{\psi})$} to denote the strength of the perturbation modes. In the limit $\epsilon_{m,n} \rightarrow 0$ we recover an integrable magnetic field lying on surfaces of constant $\rho$ with rotational transform \textcolor{black}{$\iota = \iota' \psi$} and constant shear $\iota'(\psi) = \iota'$. The perturbation is chosen to vanish at the upper \textcolor{black}{($\rho = \overline{\rho}$)} and lower ($\rho = 0$) boundaries such that there is no parallel flux into the volume. \textcolor{black}{For all of the following numerical calculations, we will take $\overline{\psi} = \overline{\rho} = L_{\zeta} = \iota' = 1$.}

\section{Transport across an island chain}
\label{sec:single_island}

In this Section, we consider the relative perpendicular and parallel transport in the presence of a single island chain generated by a resonance $\iota = n/m$. We consider the case in which $\kappa_{\perp}$ remains small but {\color{black}non-zero} such that perpendicular diffusion competes with parallel diffusion. 

{\color{black}
To analyze the behavior of the transport in such a field, in Appendix \ref{app:island_analysis} we perform a perturbation analysis in the smallness of the amplitude $\epsilon_{m,n}$. Due to the smallness of $\kappa_{\perp}$, outside of a small boundary layer of width \begin{align}
\mathcal{W}_{\text{c}} = \kappa_{\perp}^{1/4}\left(1 + \frac{n^2\overline{\rho}^2}{m^2  L_{\zeta}^2}\right)^{1/4}\left(\frac{L_{\zeta} \overline{\rho}}{m \iota' \overline{\psi}}\right)^{1/2},
\label{eq:boundary_width}
\end{align}
the perpendicular diffusion becomes unimportant. The solution is evaluated both inside and outside the boundary layer. We conclude that for both the inner and outer solution, the parallel and perpendicular diffusion can only be comparable within the island separatrices when $\mathcal{W}_{m,n}/\mathcal{W}_{\text{c}} \gtrsim 1$, where the island half-width is,
\begin{align}
    \mathcal{W}_{m,n} = \frac{2\overline{\rho}}{\overline{\psi}}\sqrt{\frac{\widetilde{\epsilon}_{m,n}(\psi)}{\iota'(\psi)}} \bigg \rvert_{\psi = n/(m\iota')} .
    \label{eq:island_half_width}
\end{align}
In the limit that $\kappa_{\perp} \rightarrow 0$, $\mathcal{W}_{m,n}/\mathcal{W}_{\text{c}} \gg 1$, and the parallel transport can compete with perpendicular transport throughout the volume within the separatrices. 
}
In this sense, an island can be considered unimportant to the transport if 
\textcolor{black}{
\begin{align}
    \mathcal{W}_{m,n} \ll \mathcal{W}_{\text{c}}.
    \label{eq:critical_width}
\end{align}
}

\begin{figure}
    \centering
    \begin{subfigure}[b]{0.48\textwidth}
    \includegraphics[width=1.0\textwidth]{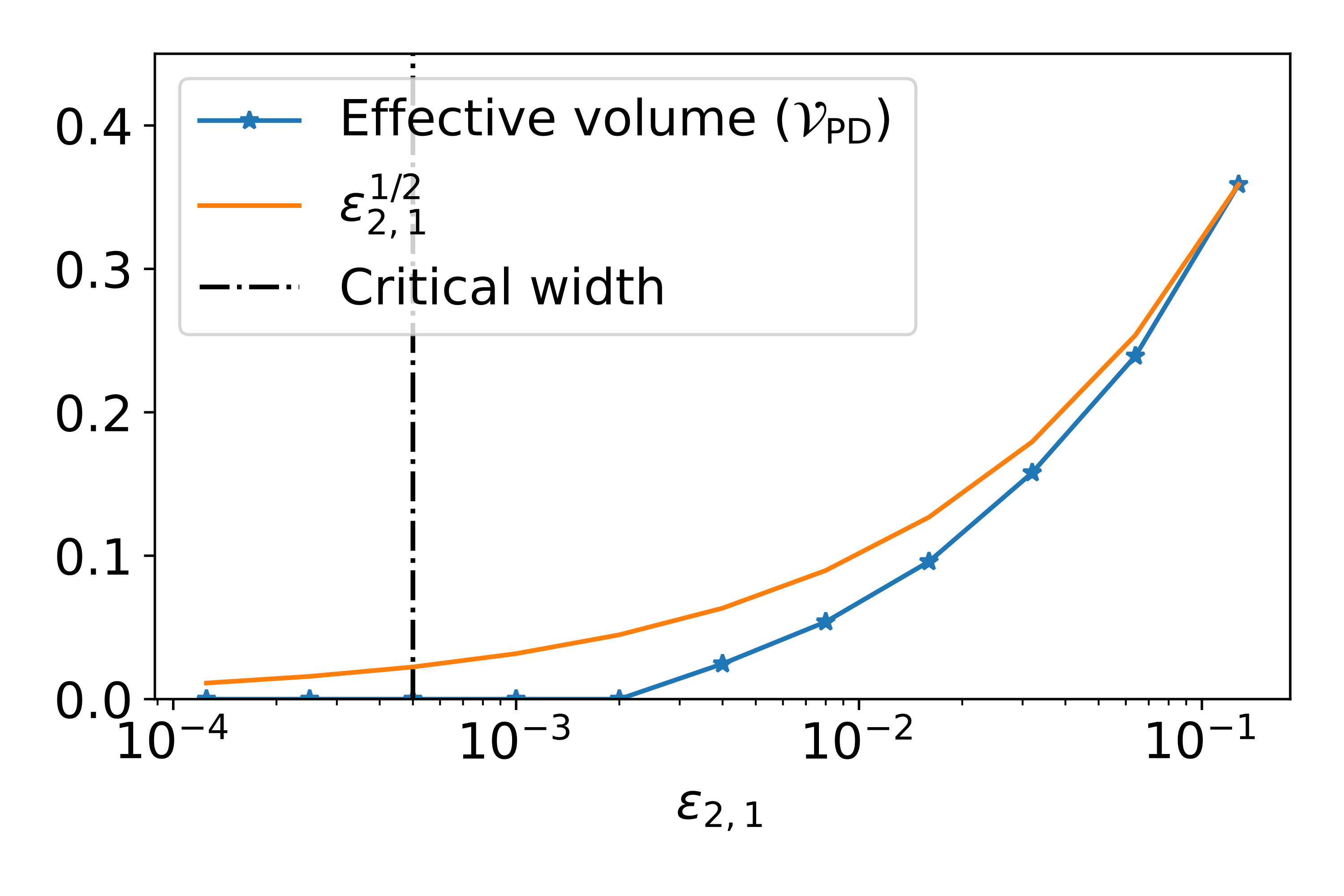}
    \caption{$\kappa_{\perp} = 10^{-6}$}
    \end{subfigure}
    \begin{subfigure}[b]{0.48\textwidth}
    \includegraphics[width=1.0\textwidth]{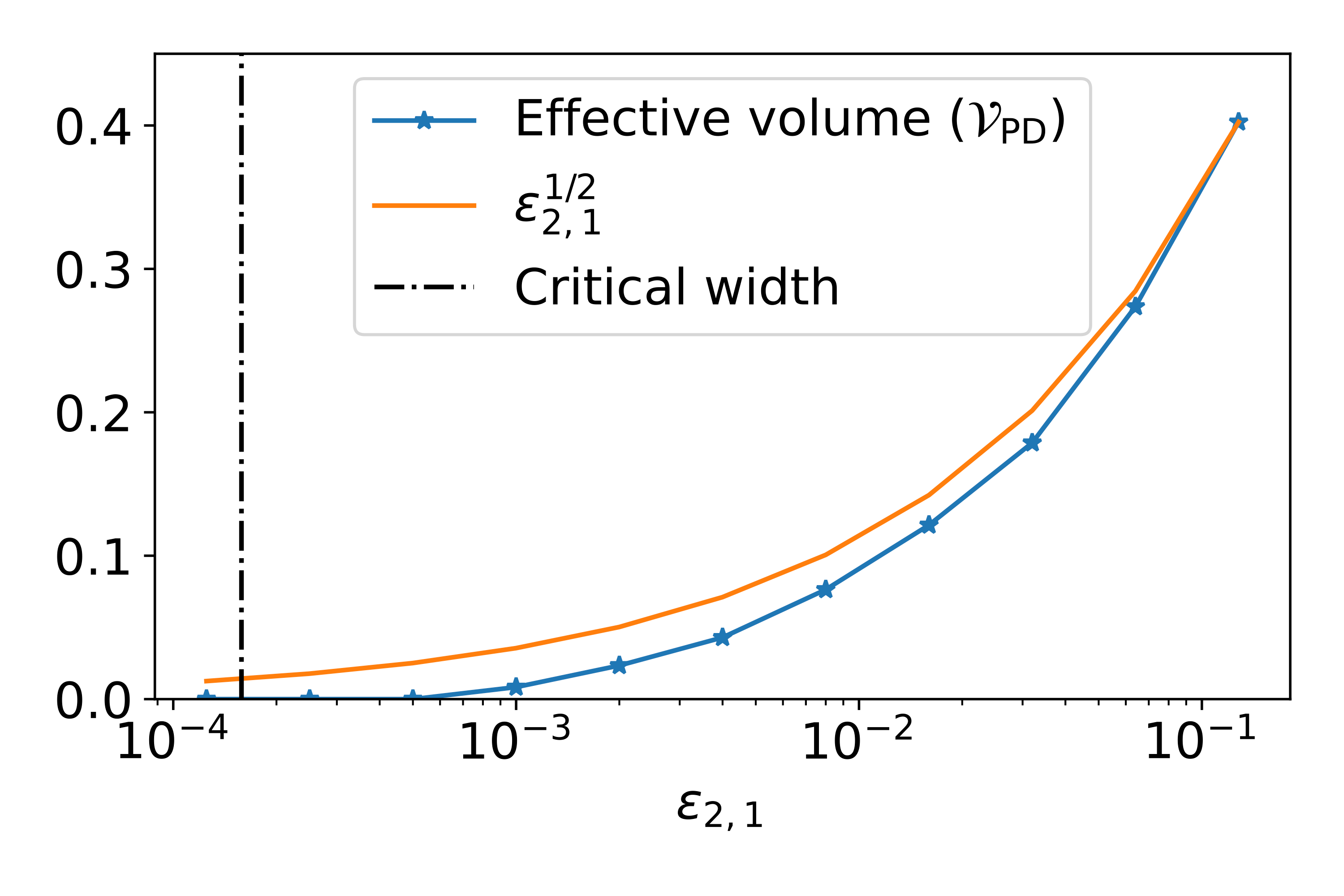}
    \caption{$\kappa_{\perp} = 10^{-7}$}
    \end{subfigure}
    \begin{subfigure}[b]{0.48\textwidth}
    \includegraphics[width=1.0\textwidth]{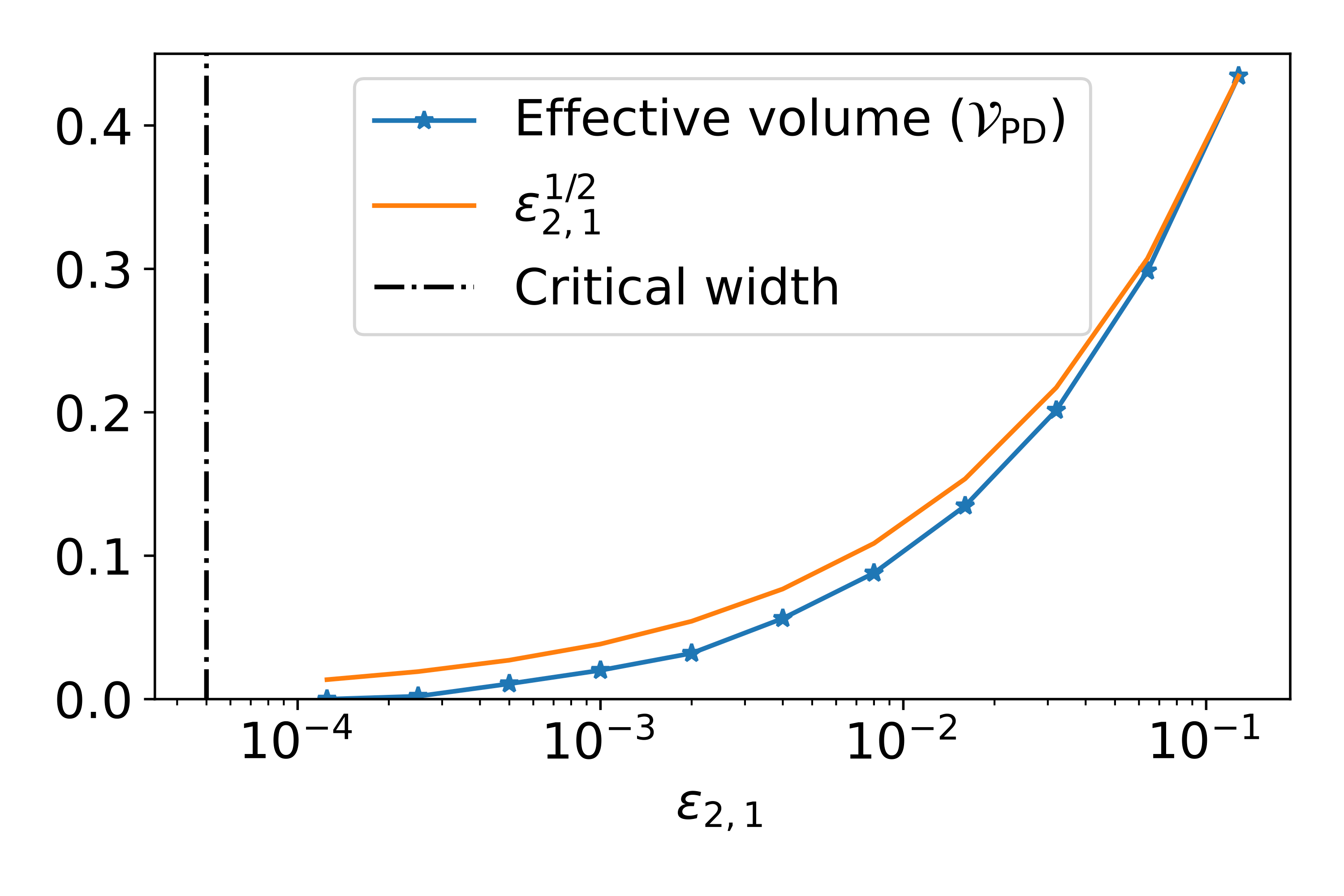}
    \caption{$\kappa_{\perp} = 10^{-8}$}
    \end{subfigure}
    \caption{The numerically computed value of the effective volume \eqref{eq:effective volume} is shown (blue) for the model field \eqref{eq:model_field} with the displayed values of $\epsilon_{2,1}$ and $\kappa_{\perp}$. Also displayed is the scaling of the island width with $\epsilon_{2,1}$ (orange), 
    and the critical value of $\epsilon_{2,1}$ for which the parallel transport may compete with perpendicular transport (black). The numerically computed value is predicted to approach the $\epsilon_{2,1}^{1/2}$ scaling in the limit $\kappa_{\perp} \rightarrow 0$.
    }
    \label{fig:single_island}
\end{figure}

In Figure \ref{fig:single_island} we present the numerically computed values of the effective volume \eqref{eq:effective volume} for a single island chain created with resonance $\epsilon_{2,1}$ with varying amplitudes and values of $\kappa_{\perp}$. The critical value, $\epsilon_{2,1}^{\text{crit}} = \sqrt{\kappa_{\perp}}/2$, predicted for the parallel transport to compete with the perpendicular transport, is indicated by the vertical dashed line. We see that, as expected, when the island is smaller than the critical width {\color{black}$\mathcal{W}_{\text{c}}$}, the computed effective volume vanishes. The point at which {\color{black}$\mathcal{V}_{\text{PD}}$} becomes positive is within an order of magnitude of $\epsilon_{2,1}^{\text{crit}}$. For large values of $\epsilon_{2,1}$, the effective volume approaches the expected scaling of the total volume enclosed by the separatrices, \textcolor{black}{$\mathcal{W}_{2,1} \sim \sqrt{\epsilon_{2,1}}$}, shown in orange. At {\color{black}non-zero} values of the perpendicular diffusion, the scaling of {\color{black}$\mathcal{V}_{\text{PD}}$} is shallower than the scaling of the total island width since the entire island chain is not dominated by parallel diffusion. 

\begin{figure}
    \centering
    \begin{subfigure}[b]{1.0\textwidth}
    \includegraphics[trim=1cm 1.05cm 1cm 1cm,clip,width=0.495\textwidth]{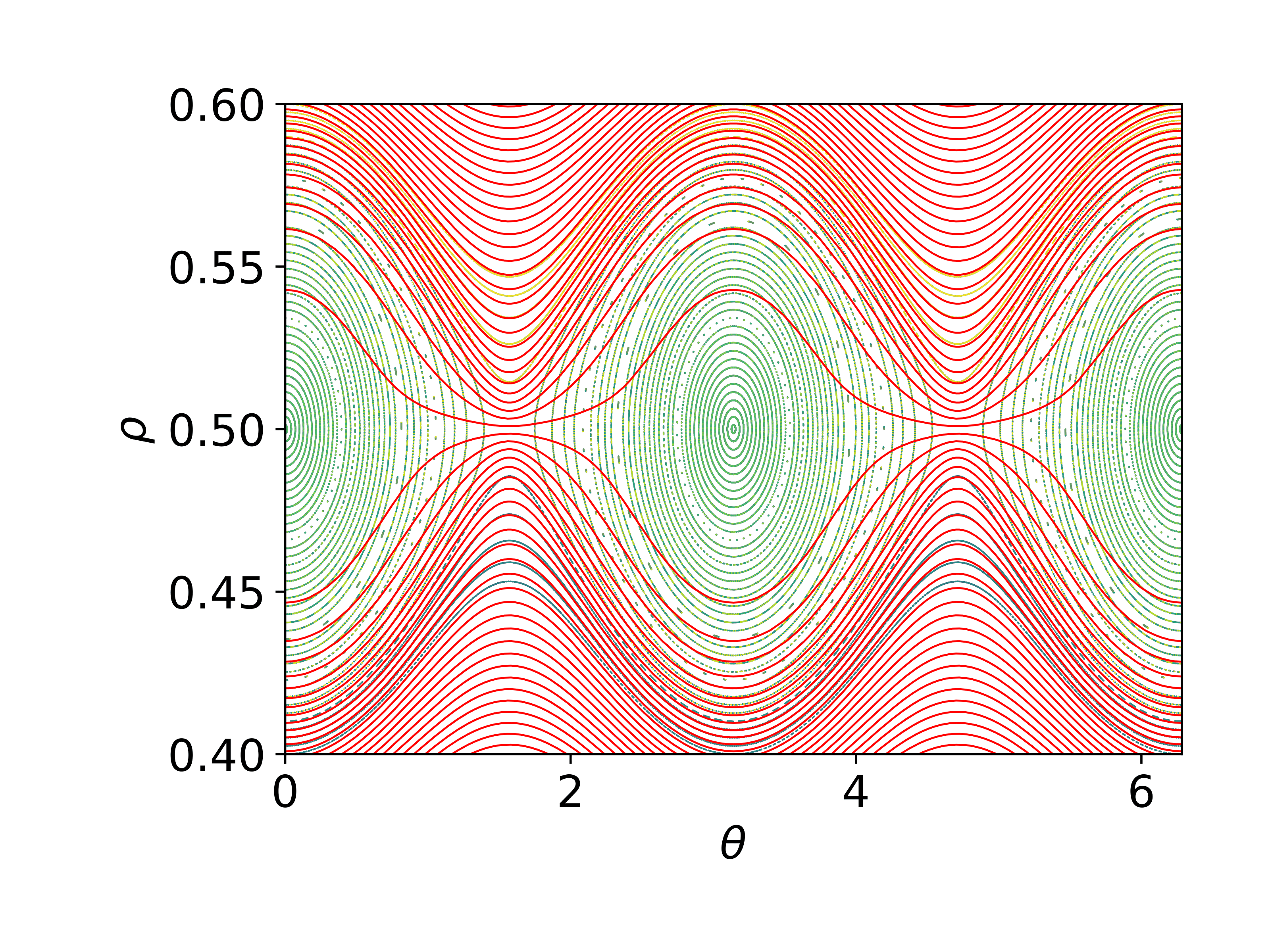}
    \includegraphics[trim=1.2cm 0.8cm 0.8cm 1cm,clip,width=0.48\textwidth]{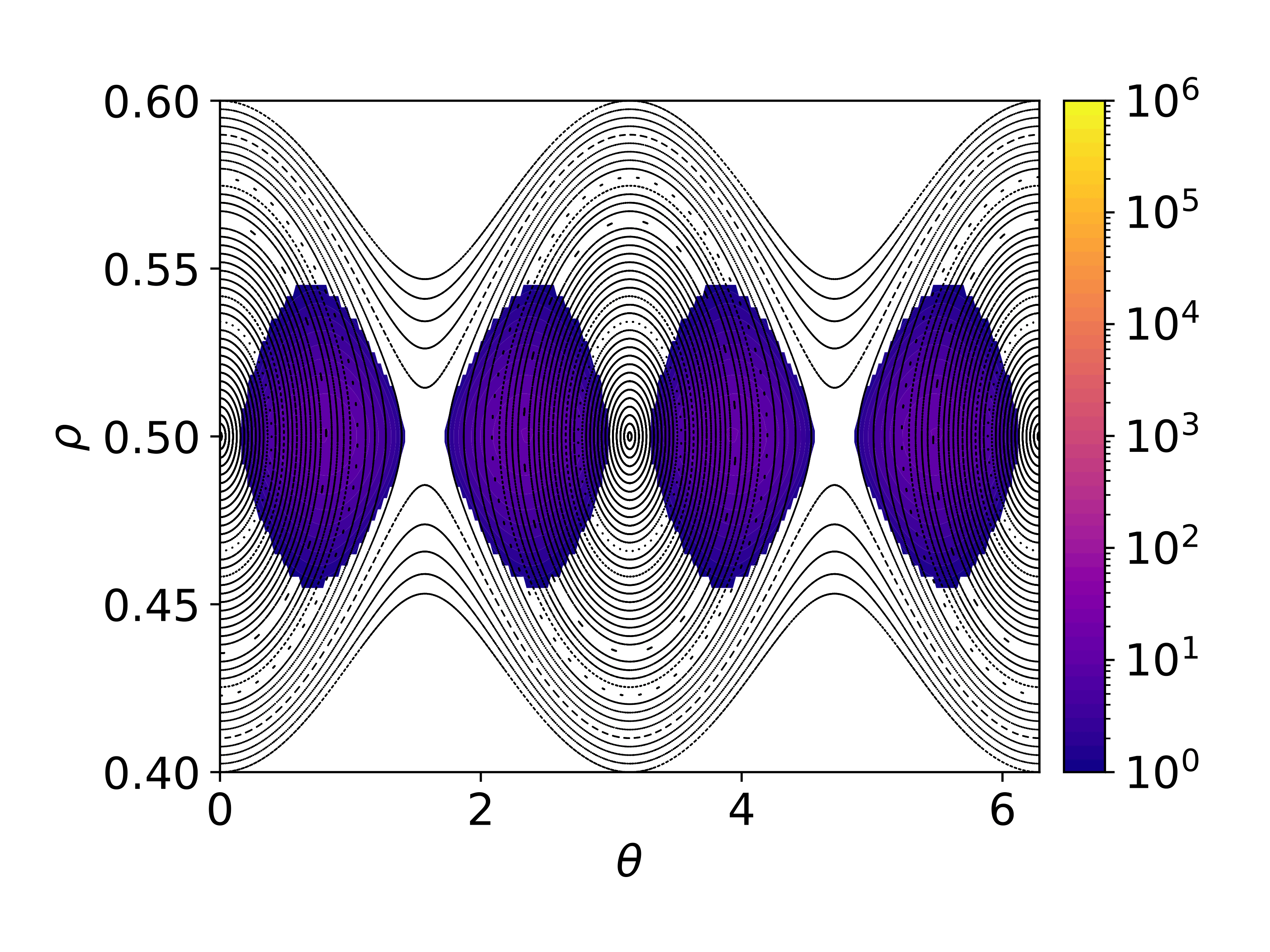}
    \caption{$\kappa_{\perp} = 10^{-6}$}
    \end{subfigure}
    \begin{subfigure}[b]{1.0\textwidth}
    \includegraphics[trim=1cm 1.05cm 1cm 1cm,clip,width=0.495\textwidth]{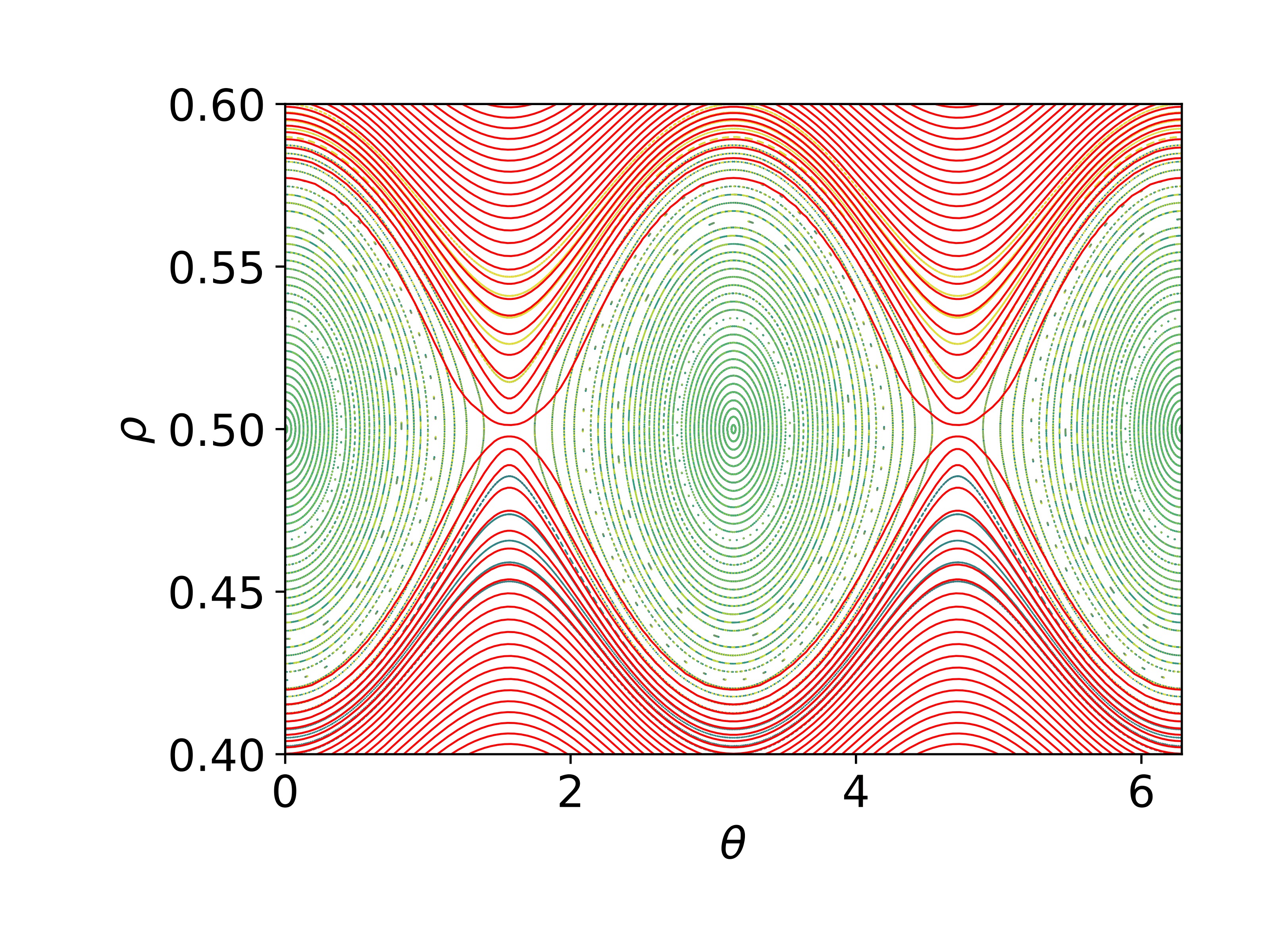}
    \includegraphics[trim=1.2cm 0.8cm 0.8cm 1cm,clip,width=0.48\textwidth]{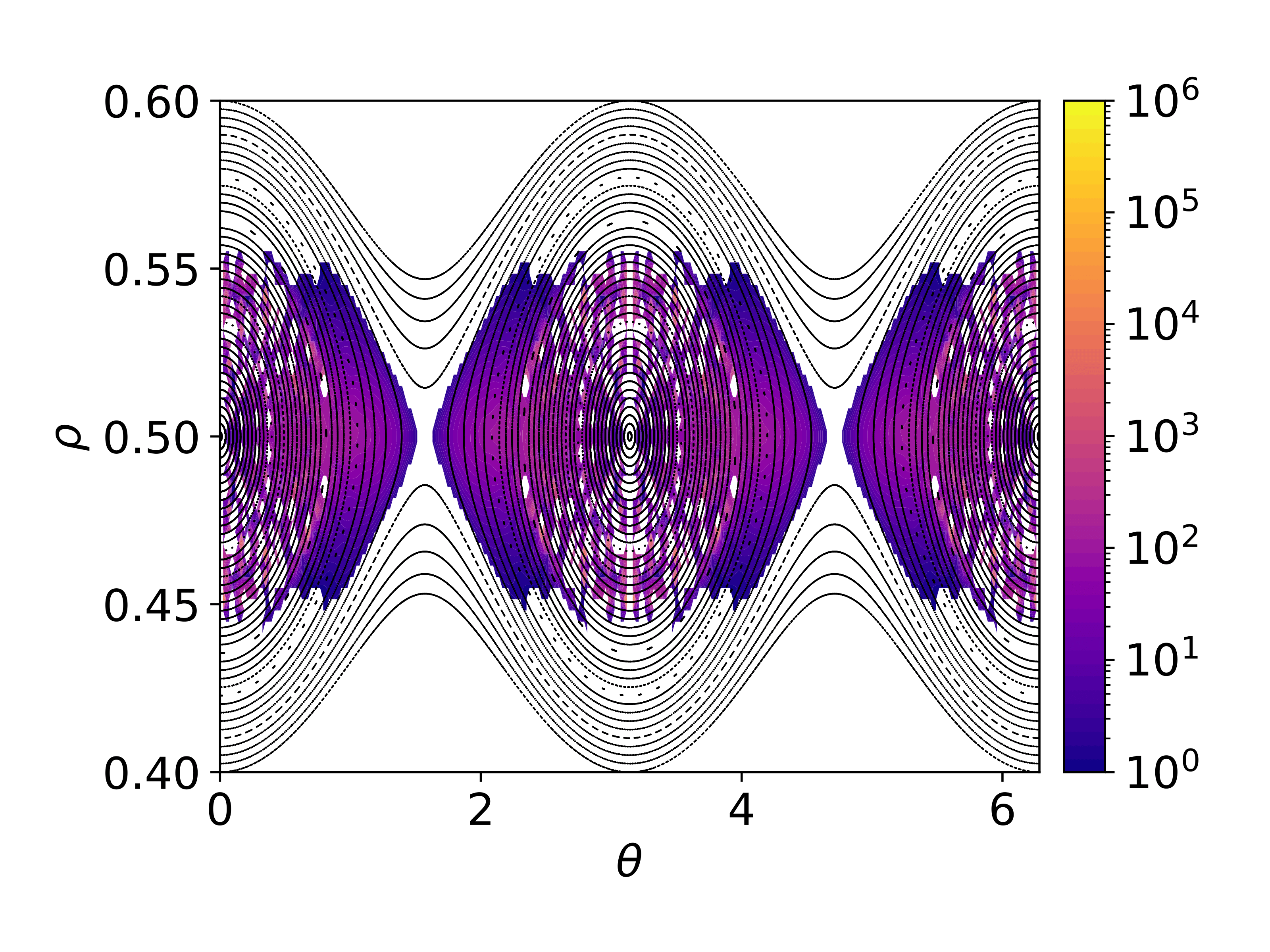}
    \caption{$\kappa_{\perp} = 10^{-7}$}
    \end{subfigure}
    \begin{subfigure}[b]{1.0\textwidth}
    \includegraphics[trim=1cm 1.05cm 1cm 1cm,clip,width=0.495\textwidth]{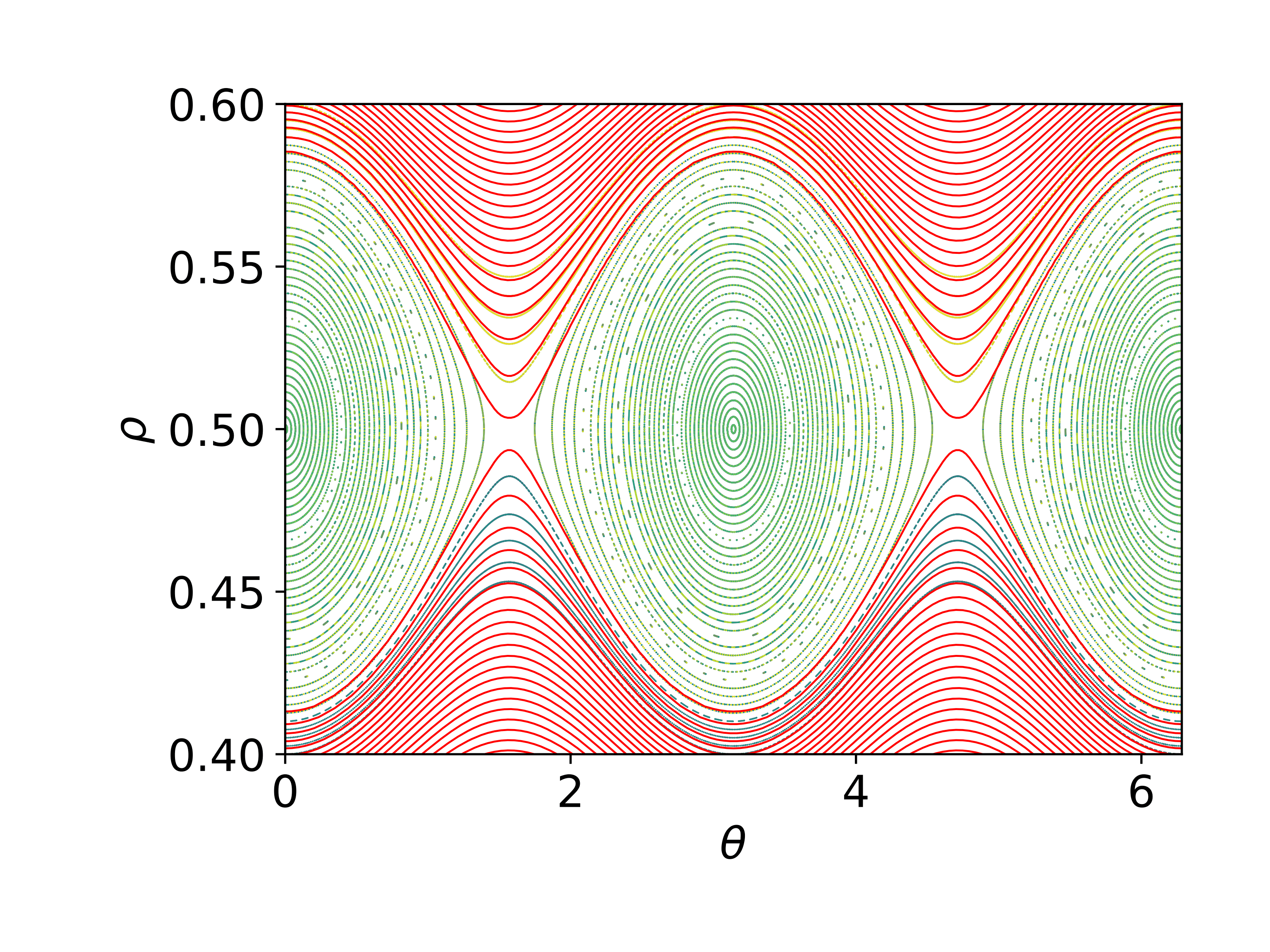}
    \includegraphics[trim=1.2cm 0.8cm 0.8cm 1cm,clip,width=0.48\textwidth]{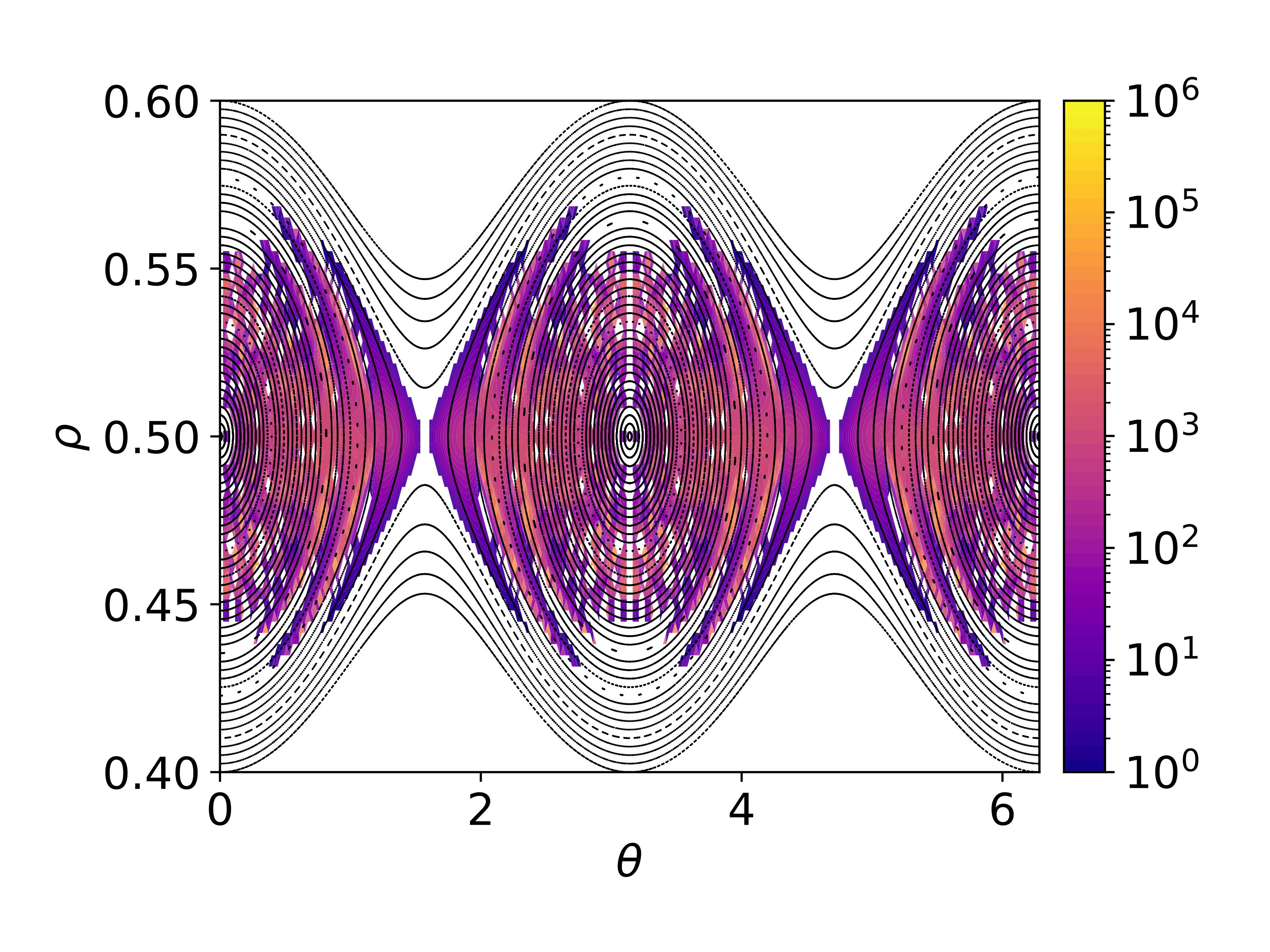}
    \caption{$\kappa_{\perp} = 10^{-8}$}
    \end{subfigure}
    \caption{The Poincar\'{e} section of the model field \eqref{eq:model_field} with $\epsilon_{2,1} = 0.008$ is displayed with isotherms (red) on the left computed with the three values of $\kappa_{\perp}$ displayed. On the right, the Poincar\'{e} section is displayed with the ratio of the parallel \eqref{eq:parallel_diffusion} to perpendicular \eqref{eq:perpendicular_diffusion} diffusion. The colorscale is {\color{black}set to white} where the parallel diffusion is smaller than the perpendicular diffusion.}
    \label{fig:diffusion_island}
\end{figure}

In Figure \ref{fig:diffusion_island} we present the isotherms (red) computed with the displayed values of $\kappa_{\perp}$ on the Poincar\'{e} section of the model magnetic field \eqref{eq:model_field} with $\epsilon_{2,1} = 0.008$. As $\kappa_{\perp}$ decreases, the isotherms conform more strongly to the structure of the magnetic field. We note the presence of isotherms which cut across the island chain at smaller values of $\kappa_{\perp}$ due to the topological restrictions.
On the right we display the Poincar\'{e} section with color highlighting the region over which the local parallel diffusion is dominating. The colorscale indicates the ratio of the local parallel to perpendicular transport. The parallel transport is strongest near the X-points, where the misalignment of the isotherms with the local field is the strongest. The parallel transport is relatively weak at the top and bottom of the island chain, where the isotherms are more strongly aligned with the magnetic field and the resonant field is weaker. As $\kappa_{\perp}$ decreases, we see the area over which the parallel transport dominates increase as well as its magnitude in comparison with the perpendicular transport.

\section{Transport across a strongly chaotic layer}
\label{sec:strong_chaos}

In this Section we consider the relative impact of perpendicular and parallel transport in the presence of strong chaos. We will assume a quasilinear model \citep{Chirikov1979,Hazeltine2003,Lichtenberg2013} to obtain a parallel diffusion coefficient. {\color{black}We will assume that the parallel transport at a given location will be dominated by nearby resonances within an island width} and that the phase of the resonances is decorrelated after one toroidal transit. We assume strong chaos such that the Chirikov island overlap parameter is large,
{\color{black}
\begin{align}
    \frac{\mathcal{W}_{m,n} + \mathcal{W}_{m',n'}}{(\Delta \rho)_{m,n;m',n'}} \gg 1,
    \label{eq:chirikov}
\end{align}
}
where $\mathcal{W}_{m,n}$ is the island half-width associated with resonances \eqref{eq:island_half_width} and \textcolor{black}{$(\Delta \rho)_{m,n;m',n'}$} is the distance between the resonances.

{\color{black}
Under the assumption that the resonances are strongly overlapping but $\epsilon_{m,n}$
 is sufficiently small that a perturbation analysis is appropriate, the motion in the field can be treated as a random walk,
 \begin{align}
    \partder{T}{\zeta} \sim D_{\|} \partder{^2T}{\psi^2},
\end{align}
with a quasi-linear diffusion coefficient given by,
\begin{align}
  D_{\|} \approx \frac{\pi}{2} \sum_{m} m^2 \left(\sum_n \Theta(\mathcal{W}_{m,n}/2 -|\rho-\rho_{\text{r}}|) \widetilde{\epsilon}_{m,n}(\psi_0) \right)^2.
  \label{eq:diffusion_coeff}
\end{align}
See Appendix \ref{app:chaos_analysis} for details.
}
The local parallel transport then scales as,
{\color{black}
\begin{align}
 |\nabla_{\|} T|^2 \sim D_{\|}^2 |\nabla_{\perp} ^2T|^2 \frac{\overline{\rho}^4}{\overline{\psi}^4  L_{\zeta}^2}.
\end{align}
We can estimate $|\nabla_{\perp}^2 T| \sim |\nabla_{\perp}T|/L_{\perp}$ for some perpendicular length scale $L_{\perp}$. Given the analysis in Section \ref{sec:single_island} and Appendix \ref{app:island_analysis}, we can note several relevant length scales. Within the boundary layer, $L_{\perp} \sim \mathcal{W}_{\text{c}}$. Outside a width $\mathcal{W}_{m,n}$ of a given resonance and outside the boundary layer, $L_{\perp} \sim \overline{\rho}$.  Within a width $\mathcal{W}_{m,n}$ and outside the boundary layer, $L_{\perp} \sim |\rho-\rho_{\text{r}}|$. Given that the resonances are strongly overlapping, the perpendicular temperature gradient will be dominated by nearby resonances. Thus we can estimate $L_{\perp} \lesssim \mathcal{W}_{m,n}$ and,
\begin{align}
   |\nabla_{\|} T|^2\gtrsim D_{\|}^2 \frac{|\nabla_{\perp} T|^2}{\mathcal{W}_{m,n}^2} \frac{\overline{\rho}^4}{\overline{\psi}^4L_{\zeta}^2}.
\end{align}}
We conclude that the parallel transport may compete with the perpendicular transport when,
{\color{black}
\begin{align}
    \frac{D_{\|}^2 \overline{\rho}^4}{\mathcal{W}_{m,n}^2 \overline{\psi}^4L_{\zeta}^2} \sim \kappa_{\perp}.
    \label{eq:critical_kappaperp}
\end{align}
}

\begin{figure}
    \centering
    \includegraphics[width=0.5\textwidth]{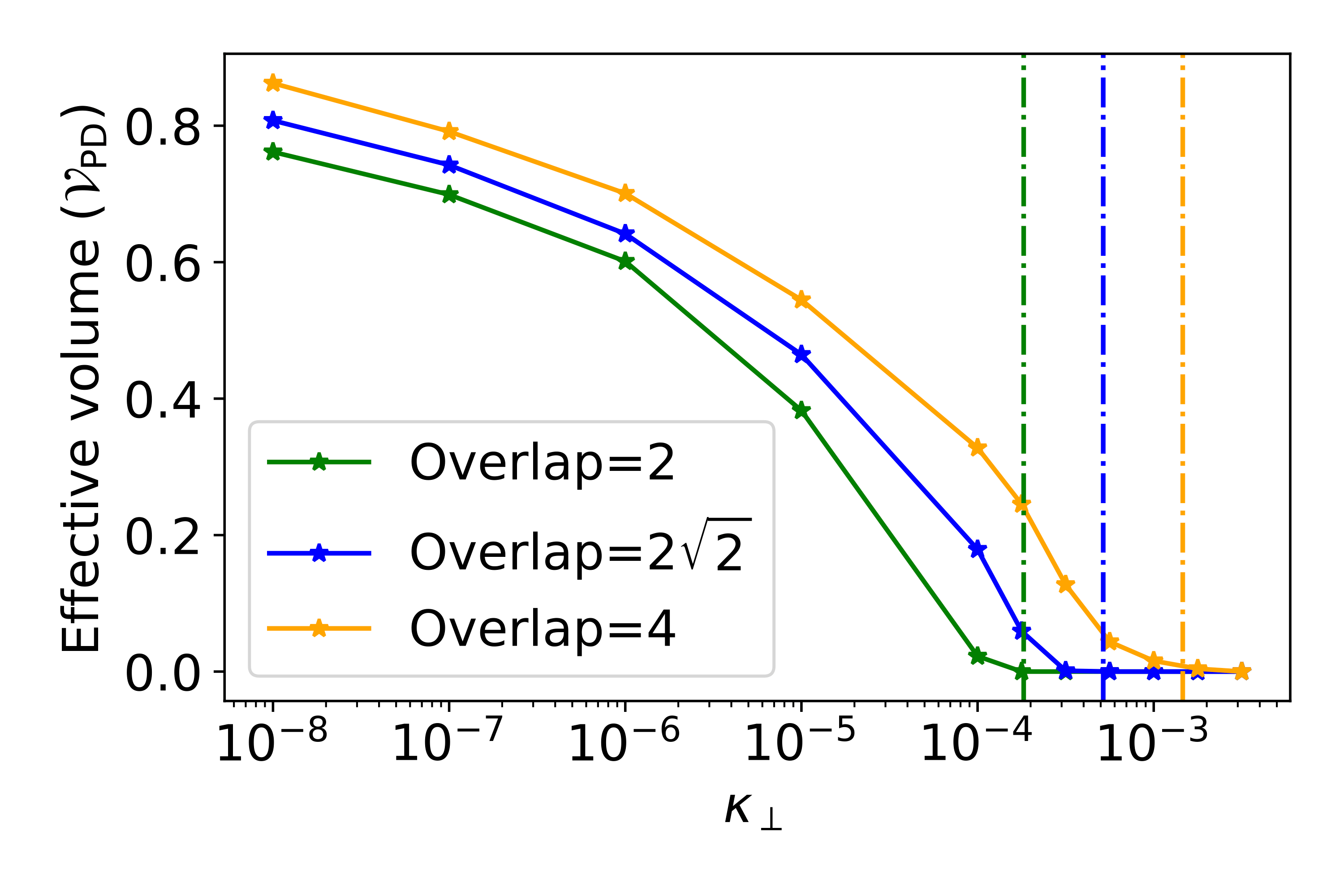}
    \caption{The numerically computed value of the effective volume \eqref{eq:effective volume} is shown for the model field \eqref{eq:model_field} with $m=12$ resonances with amplitudes chosen to provide strong island overlap (values of the overlap parameter \eqref{eq:chirikov} are indicated). Also displayed is the critical value of $\kappa_{\perp}$ for which the parallel transport is predicted to compete with perpendicular transport (vertical dashed). 
    }
    \label{fig:diffusion_strong_scaling}
\end{figure}

To test the expected scaling, we consider the model field \eqref{eq:model_field} with $m = 12$ and $n = [2,\dots,10]$. We evaluate three strongly chaotic fields with amplitudes $\epsilon_{m,n}$ chosen such that the Chirikov parameter \eqref{eq:chirikov} is 2, $2 \sqrt{2}$, and 4. We compute solutions to the ADE for a range of $\kappa_{\perp}$. The effective volume \eqref{eq:effective volume} is shown in Figure \ref{fig:diffusion_strong_scaling} along with the critical value of $\kappa_{\perp}$ (vertical dashed) at which the parallel diffusion may compete \eqref{eq:critical_kappaperp}. We note that the point at which the effective volume becomes non-zero is within an order of magnitude of that expected from quasilinear theory for each value of the overlap parameter. As $\kappa_{\perp}$ decreases, the effective volume computed from the three model fields approach each other. 

\begin{figure}
    \centering
    \begin{subfigure}[b]{1.0\textwidth}
    \includegraphics[trim=1cm 1.05cm 1cm 1cm,clip,width=0.495\textwidth]{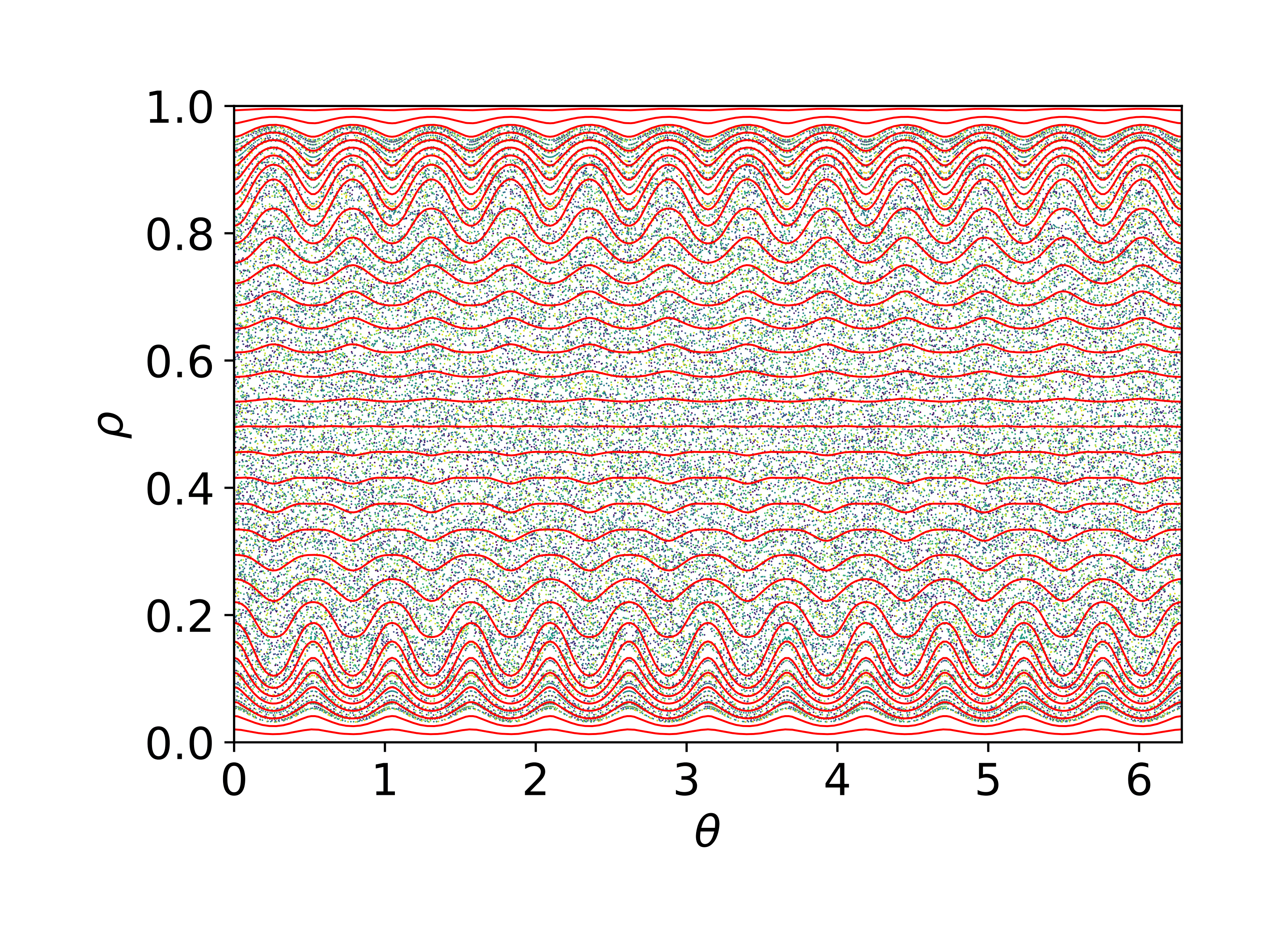}
    \includegraphics[trim=1.2cm 0.8cm 0.8cm 1cm,clip,width=0.48\textwidth]{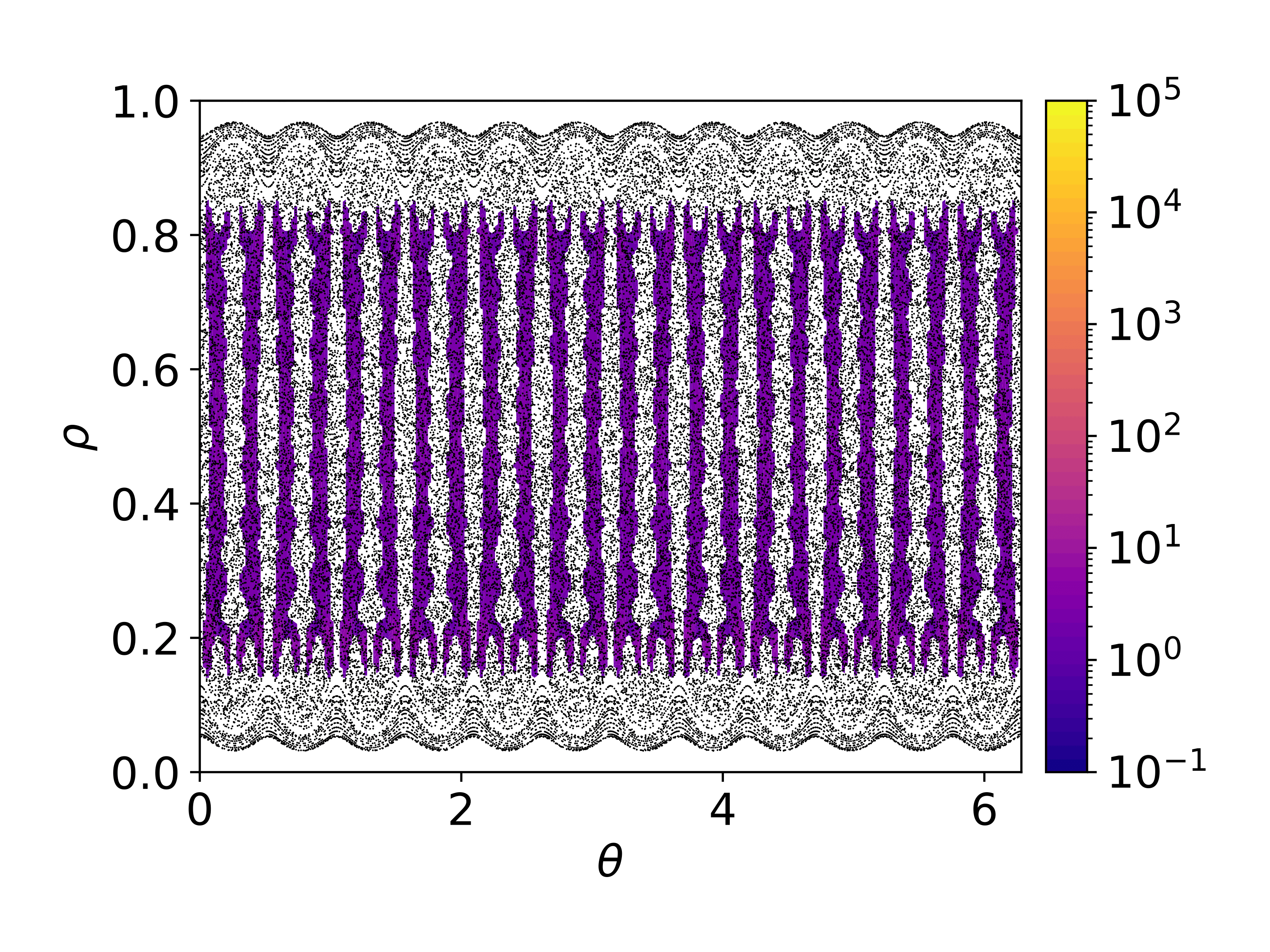}
    \caption{$\kappa_{\perp} = 10^{-4}$}
    \end{subfigure}
    \begin{subfigure}[b]{1.0\textwidth}
    \includegraphics[trim=1cm 1.05cm 1cm 1cm,clip,width=0.495\textwidth]{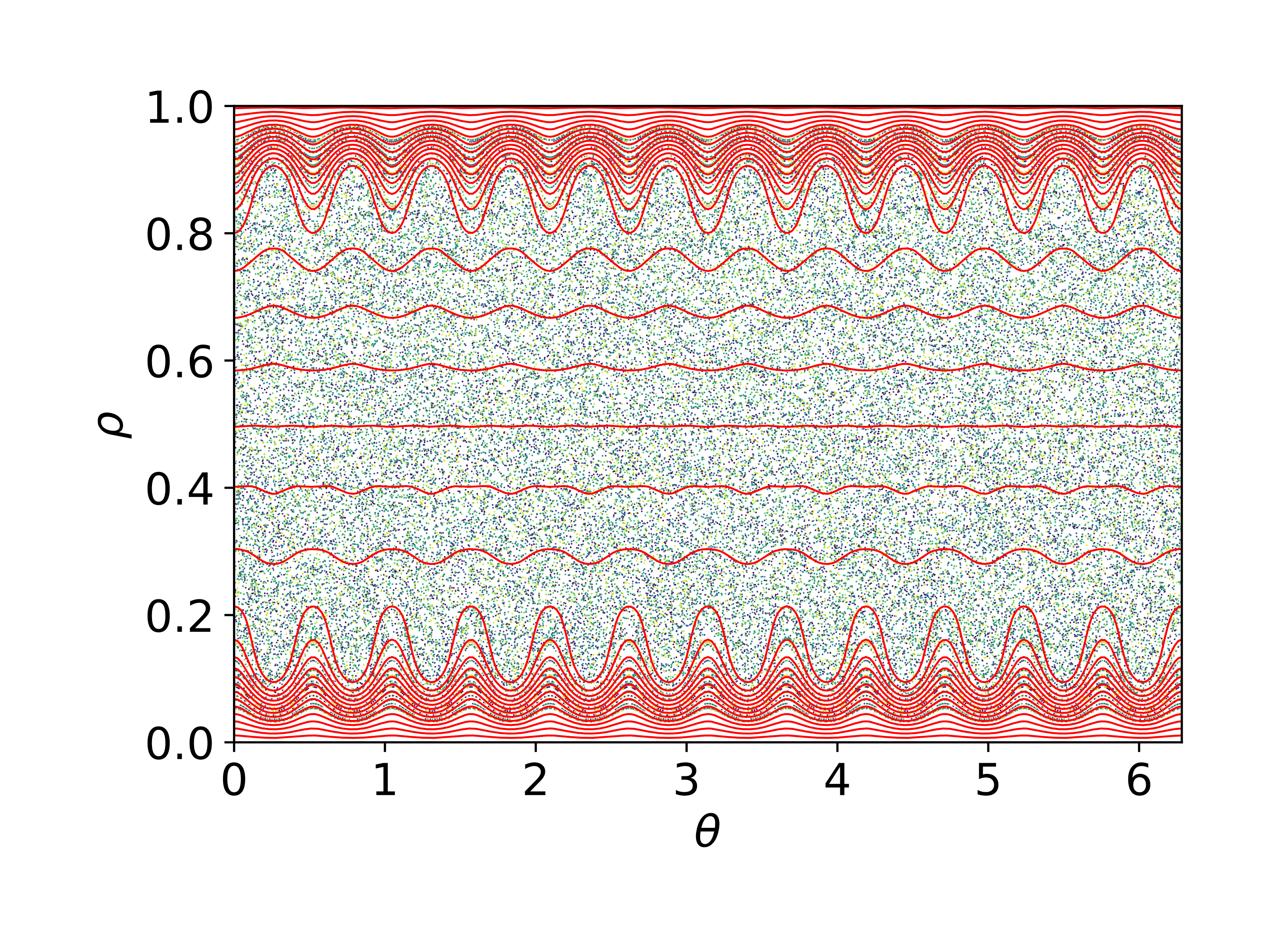}
    \includegraphics[trim=1.2cm 0.8cm 0.8cm 1cm,clip,width=0.48\textwidth]{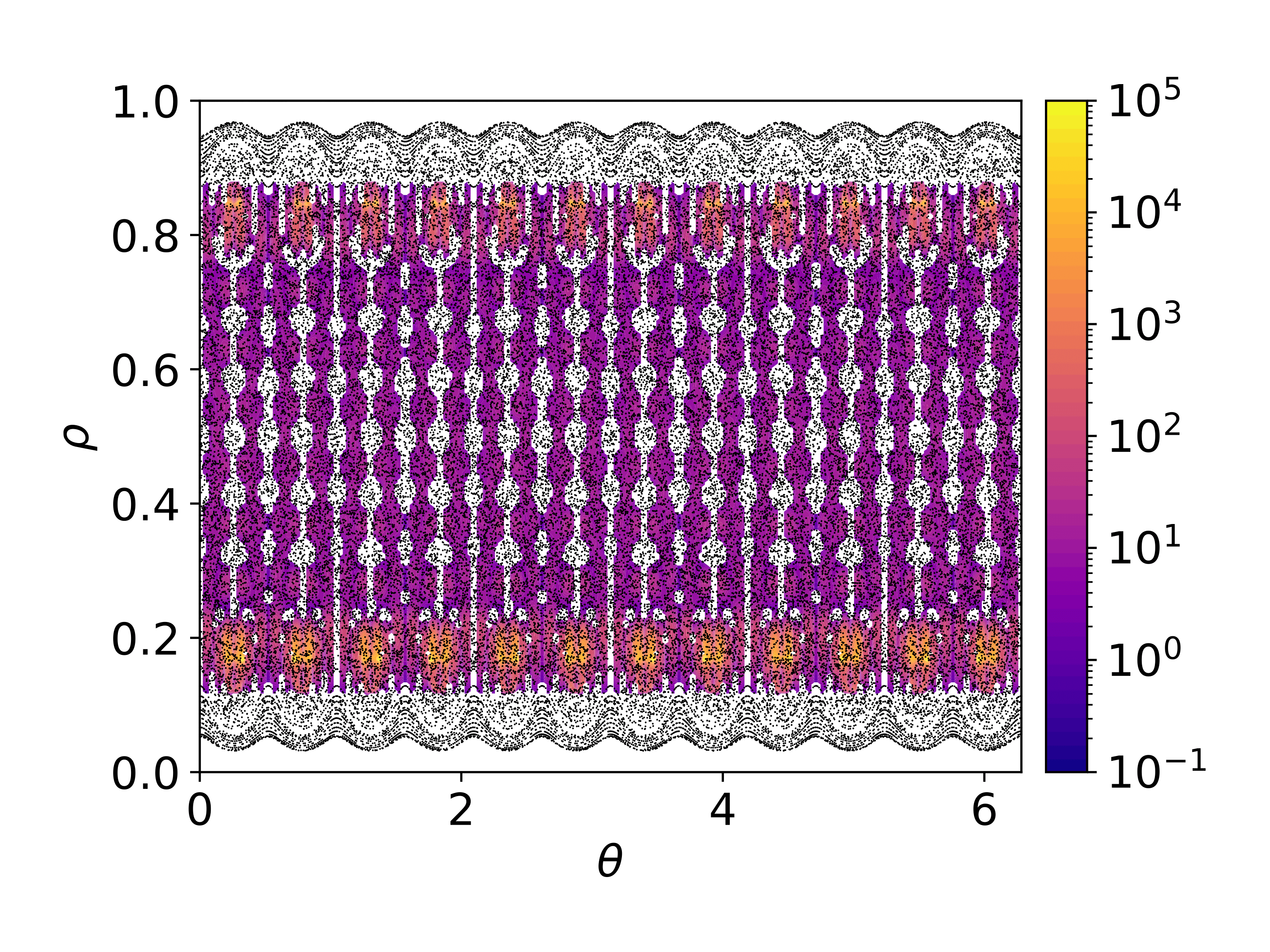}
    \caption{$\kappa_{\perp} = 10^{-5}$}
    \end{subfigure}
    \begin{subfigure}[b]{1.0\textwidth}
    \includegraphics[trim=1cm 1.05cm 1cm 1cm,clip,width=0.495\textwidth]{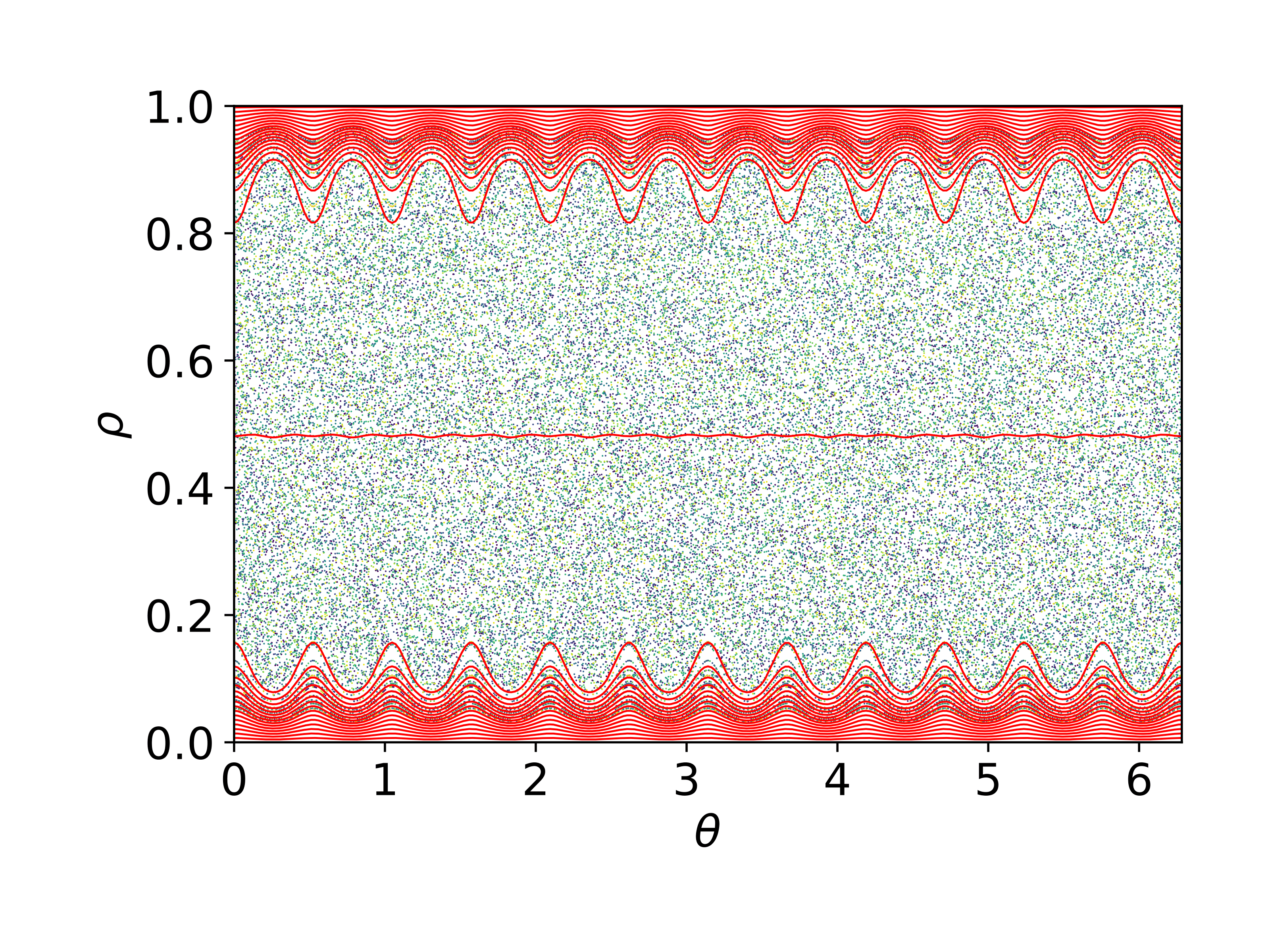}
    \includegraphics[trim=1.2cm 0.8cm 0.8cm 1cm,clip,width=0.48\textwidth]{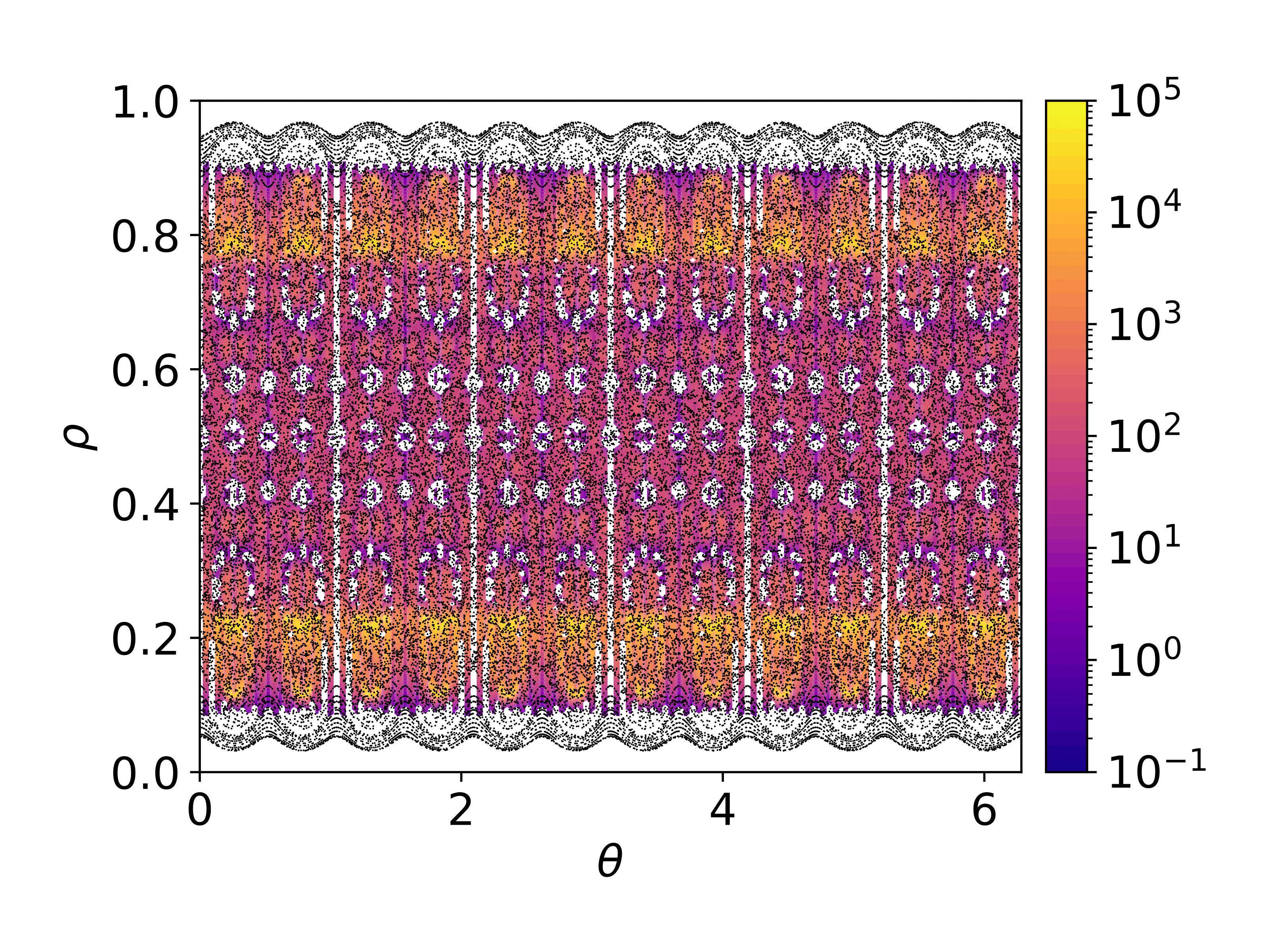}
    \caption{$\kappa_{\perp} = 10^{-6}$}
    \end{subfigure}
    \caption{The Poincar\'{e} section of the model field \eqref{eq:model_field} with $m = 12$ resonances chosen to provide an overlap parameter \eqref{eq:chirikov} of 4 is displayed with isotherms (red) on the left for the three values of $\kappa_{\perp}$ displayed. On the right, the Poincar\'{e} section is displayed with the ratio of the parallel \eqref{eq:parallel_diffusion} to perpendicular \eqref{eq:perpendicular_diffusion} diffusion. The colorscale is {\color{black}set to white} where the parallel diffusion is smaller than the perpendicular diffusion.}
    \label{fig:strong_chaos}
\end{figure}

In Figure \ref{fig:strong_chaos} we display the Poincar\'{e} section with isotherms (left) and with the parallel-transport dominated regions highlighted in color (right). Although the field is strongly chaotic, we note several isotherms near the boundaries which adapt to the $m = 12$ structure of the residual island chains. For $\kappa_{\perp} = 10^{-4}$, we note only a very small volume over which the parallel transport dominates. As we decrease the value of $\kappa_{\perp}$, we see that the parallel transport dominates over much of the volume contained within the outermost island separatrices and the temperature gradient across the chaotic layer is significantly reduced. We note an $n = 6$ variation in the structure of the parallel diffusion due to the poloidal dependence of the radial field \eqref{eq:model_field}. Near the edges of the chaotic layer, the isotherms become strongly shaped in comparison with the relatively flat isotherms in the chaotic layer. Given that the magnitude of the radial field is smaller than the toroidal field by the amplitude of the perturbation, this feature enhances the parallel diffusion near the boundaries of the chaotic layer.

\section{Transport across a critically chaotic layer}
\label{sec:critical_chaos}

Although simple analytic scalings can be derived in the case of a single island chain or strong chaos, in reality a magnetic field may possess a complicated mixture of KAM surfaces, partial barriers, chaos, and island chains. In this case, we can use the metric defined in \S \ref{sec:effective_volume} to assess the {\color{black} relative impact of non-integrability on the transport through the volume of parallel diffusion}. 

As a point of comparison, we will consider four marginally chaotic magnetic fields generated by resonances with different poloidal mode numbers. The amplitude of the perturbation will be chosen to achieve critical island overlap with a comparable volume enclosed by the outermost separatrices. For the first field, we choose $n/m = [1/4, 2/4, 3/4]$. Since the resonances are equally spaced with $\Delta \psi = 1/4$, we choose the perturbation amplitudes $\epsilon_{m,n}$ such that the associated half island widths \eqref{eq:island_half_width} are $1/8$. We similarly consider fields with $n/m =  [2/12,\dots,10/12]$, $n/m =[3/20,\dots,17/20]$, and $n/m = [5/36,\dots,31/36]$ with critical island overlap. The separatrices associated with the outermost island chains for each of these model fields are located at $\rho =1/8$ and $\rho = 7/8$ such that they have a comparable ``volume'' of chaos. 



\begin{figure}
    \centering
    \begin{subfigure}[b]{0.48\textwidth}
    \includegraphics[width=1.0\textwidth]{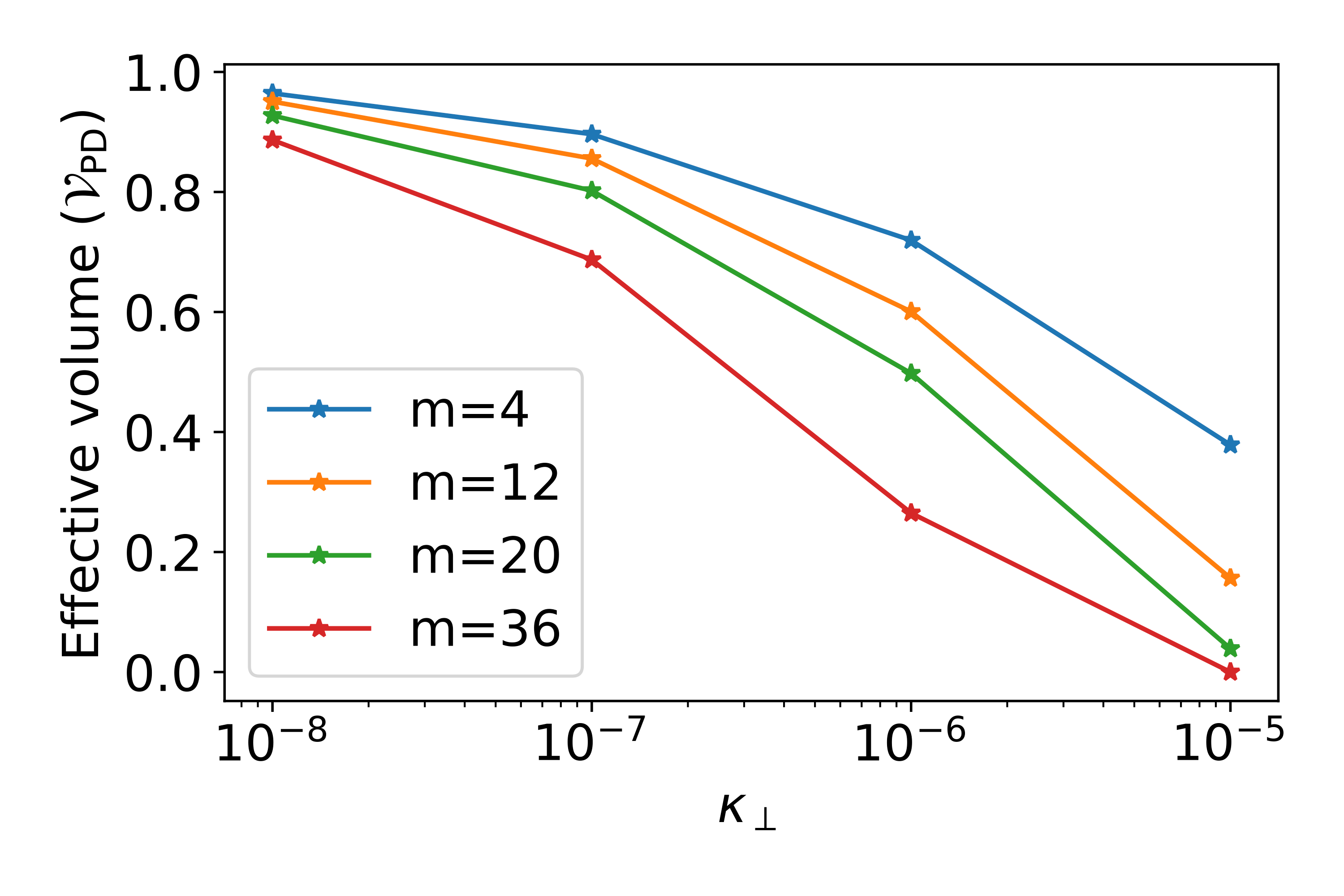}
    \caption{}
    \end{subfigure}
    \begin{subfigure}[b]{0.48\textwidth}
    \includegraphics[width=1.0\textwidth]{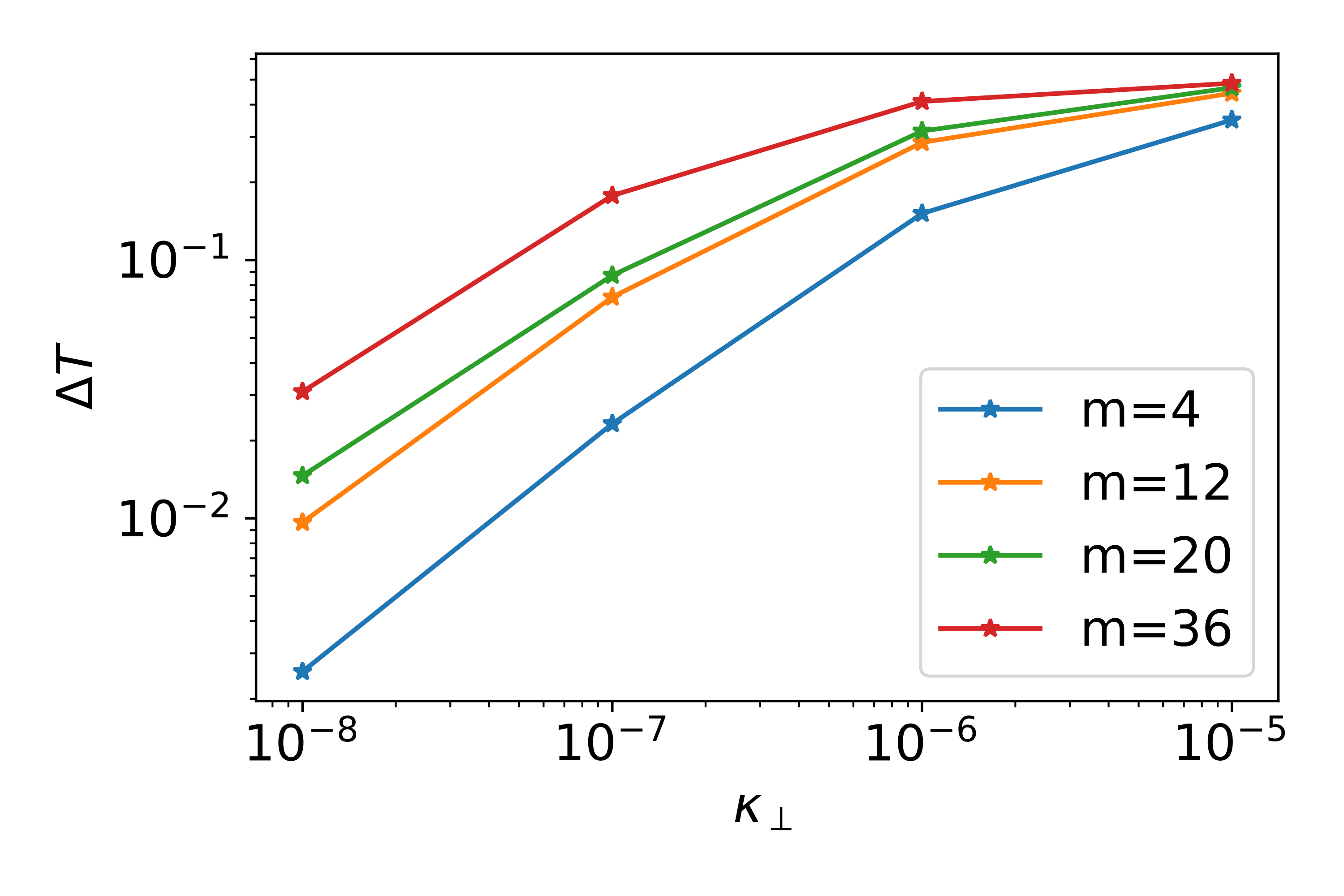}
    \caption{}
    \end{subfigure}
    \caption{(a) The effective volume \eqref{eq:effective volume} is computed for the model field \eqref{eq:model_field} with the displayed mode number perturbations with amplitudes $\epsilon_{m,n}$ chosen for critical island overlap. The effective volume is computed over a subset of the entire volume, $\rho \in [0.25, 0.75]$. 
    (b) The total temperature difference between $\rho = 0.75$ and $\rho = 0.25$ upon averaging over the angles.}
    \label{fig:critical_chaos_scaling}
\end{figure}

If the remnants of the island chains are dominating the transport, we anticipate the parallel transport to dominate when,
\begin{align}
    \epsilon_{m,n}^2 m^2  \sim \kappa_{\perp},
\end{align}
from \eqref{eq:critical_width}, while if strong chaos is dominating the transport, 
\begin{align}
    \epsilon_{m,n}^3 m^4 \sim \kappa_{\perp},
\end{align}
from \eqref{eq:critical_kappaperp}. Since we have chosen $\epsilon_{m,n} \sim 1/m^2$ to achieve critical overlap, we expect the parallel transport to be relatively more impactful when $m$ is small. In Figure \ref{fig:critical_chaos_scaling} we compare the effective volume between the model fields. Since the volume contained within the outermost separatrices differs slightly between the two model fields, we compute the effective volume only for $\rho \in [0.25, 0.75]$. As anticipated, the $m = 4$ resonances produce a larger effective volume of {\color{black}parallel diffusion}, and the relative difference between the model fields increases with increasing $\kappa_{\perp}$. To quantify the effect of the field structure on temperature flattening, we also compare the total temperature differential between $\rho = 0.75$ and $\rho = 0.25$. At large values of $\kappa_{\perp}$, each of the model fields can support a significant temperature gradient, although it is reduced for the $m =4$ field.

\begin{figure}
    \begin{subfigure}[b]{1.0\textwidth}
    \includegraphics[trim=1cm 1.05cm 1cm 1cm,clip,width=0.495\textwidth]{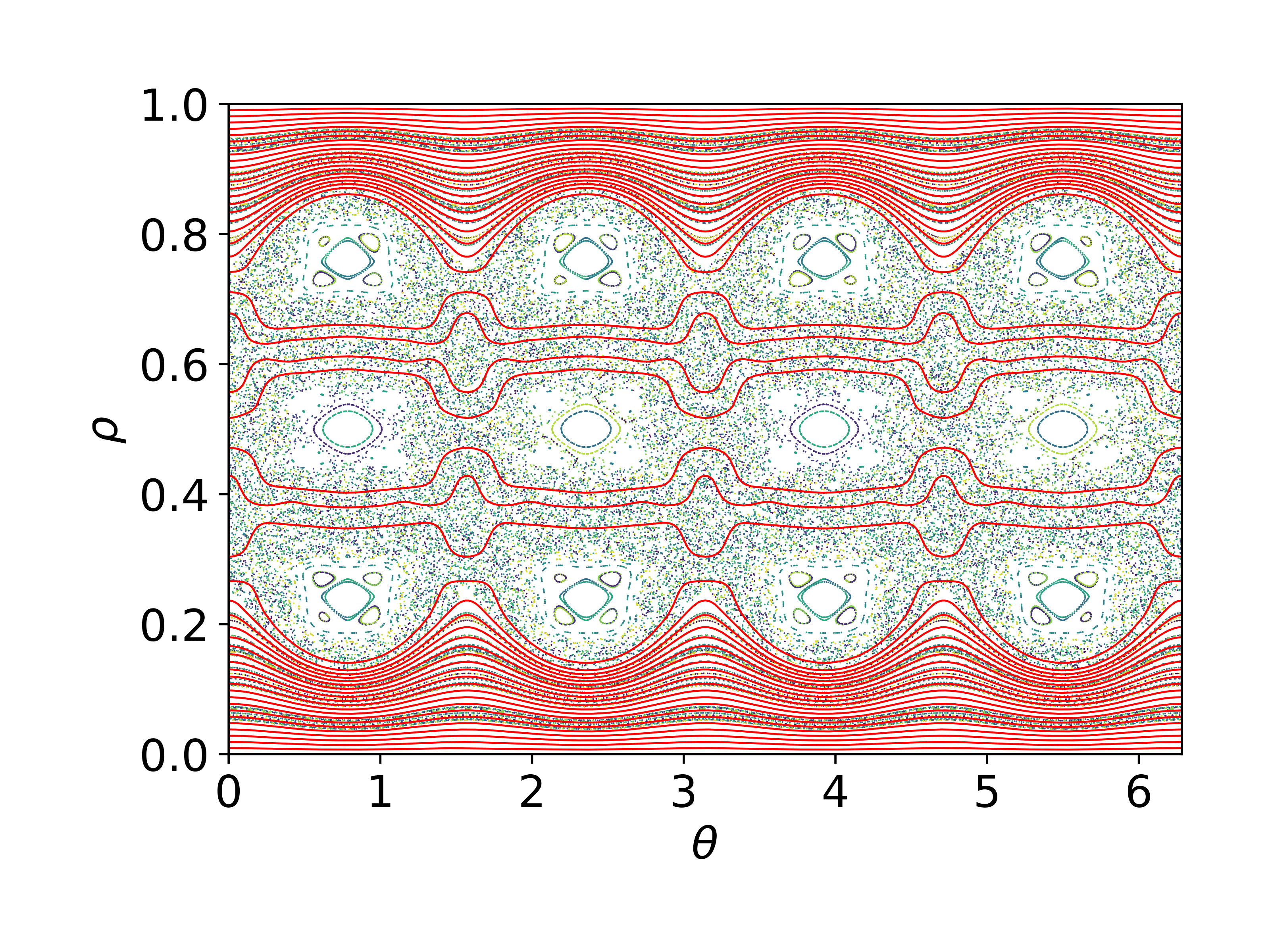}
    \includegraphics[trim=1.2cm 0.8cm 0.8cm 1cm,clip,width=0.48\textwidth]{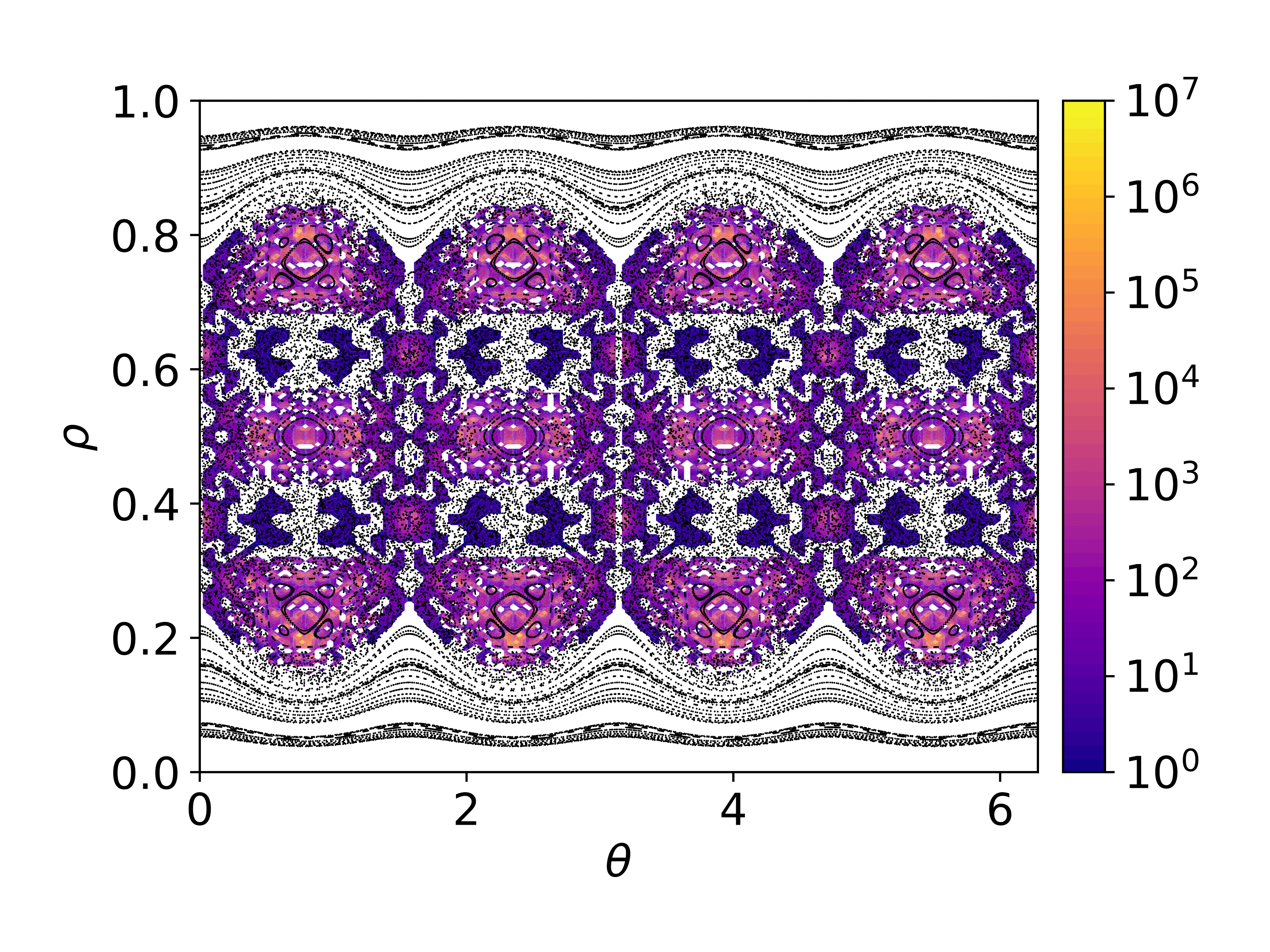}
    \caption{$m=4$}
    \end{subfigure}
    \begin{subfigure}[b]{1.0\textwidth}
    \includegraphics[trim=1cm 1.05cm 1cm 1cm,clip,width=0.495\textwidth]{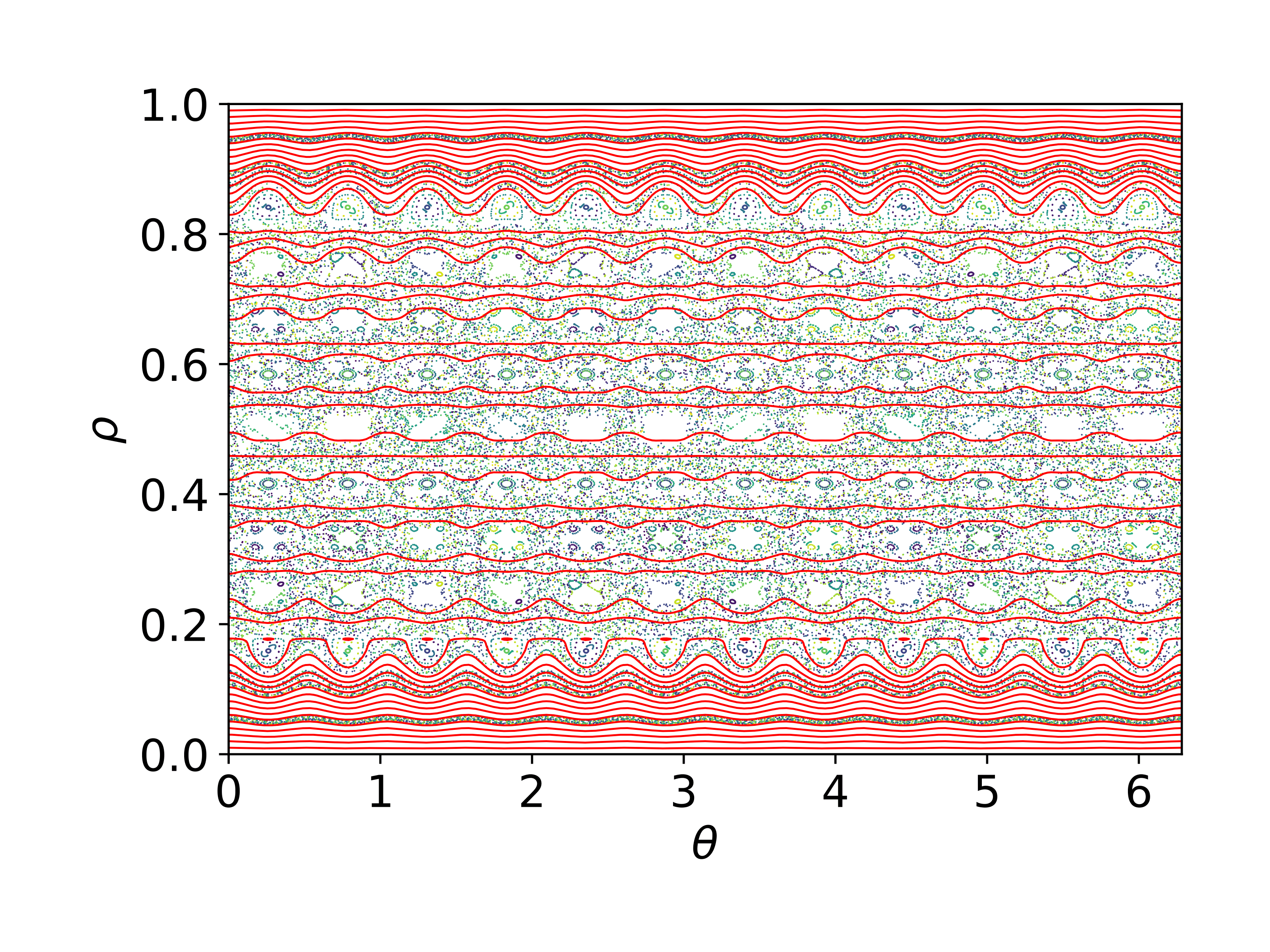}
    \includegraphics[trim=1.2cm 0.8cm 0.8cm 1cm,clip,width=0.48\textwidth]{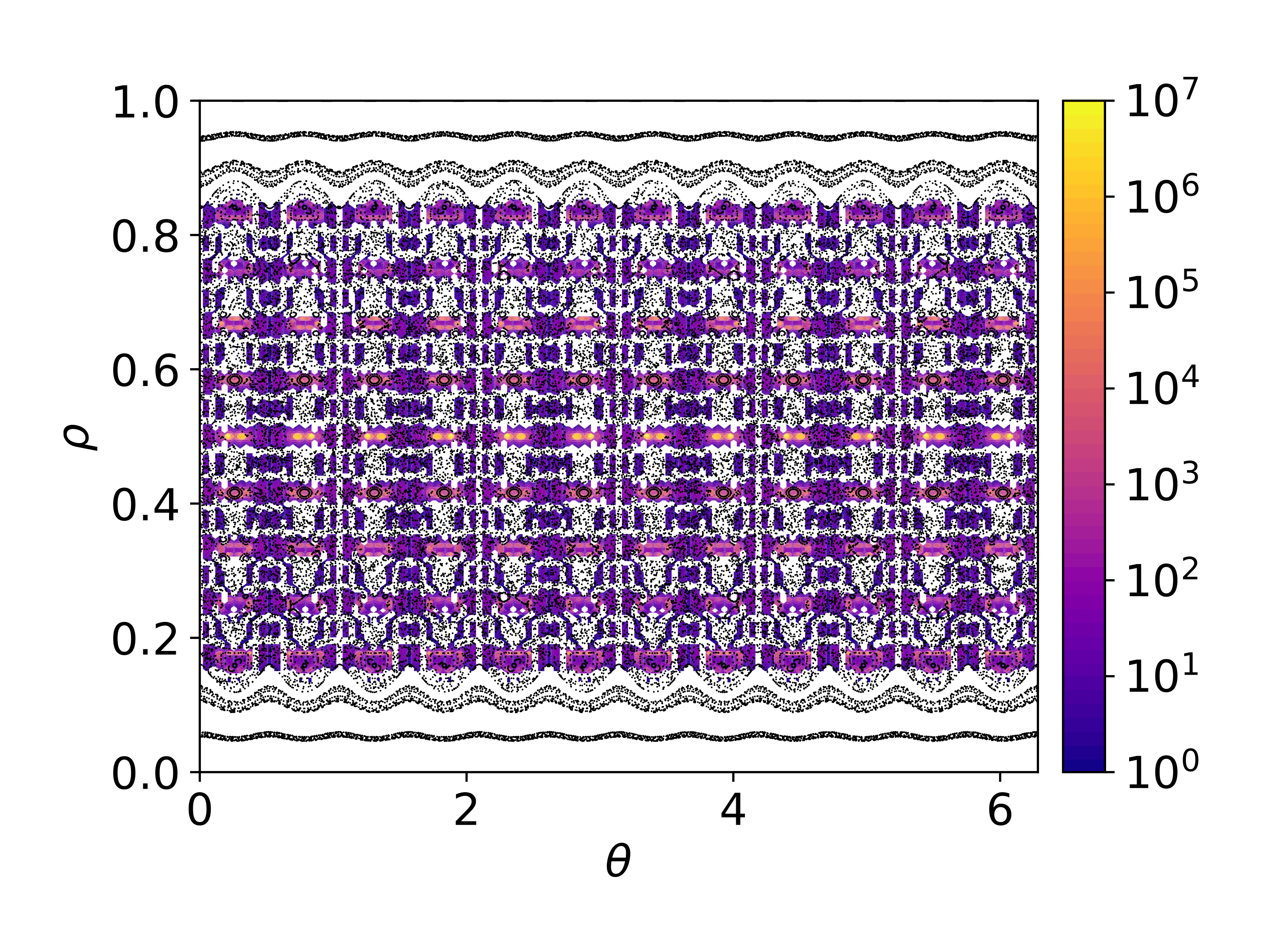}
    \caption{$m = 12$}
    \end{subfigure}
    \begin{subfigure}[b]{1.0\textwidth}
    \includegraphics[trim=1cm 1.05cm 1cm 1cm,clip,width=0.495\textwidth]{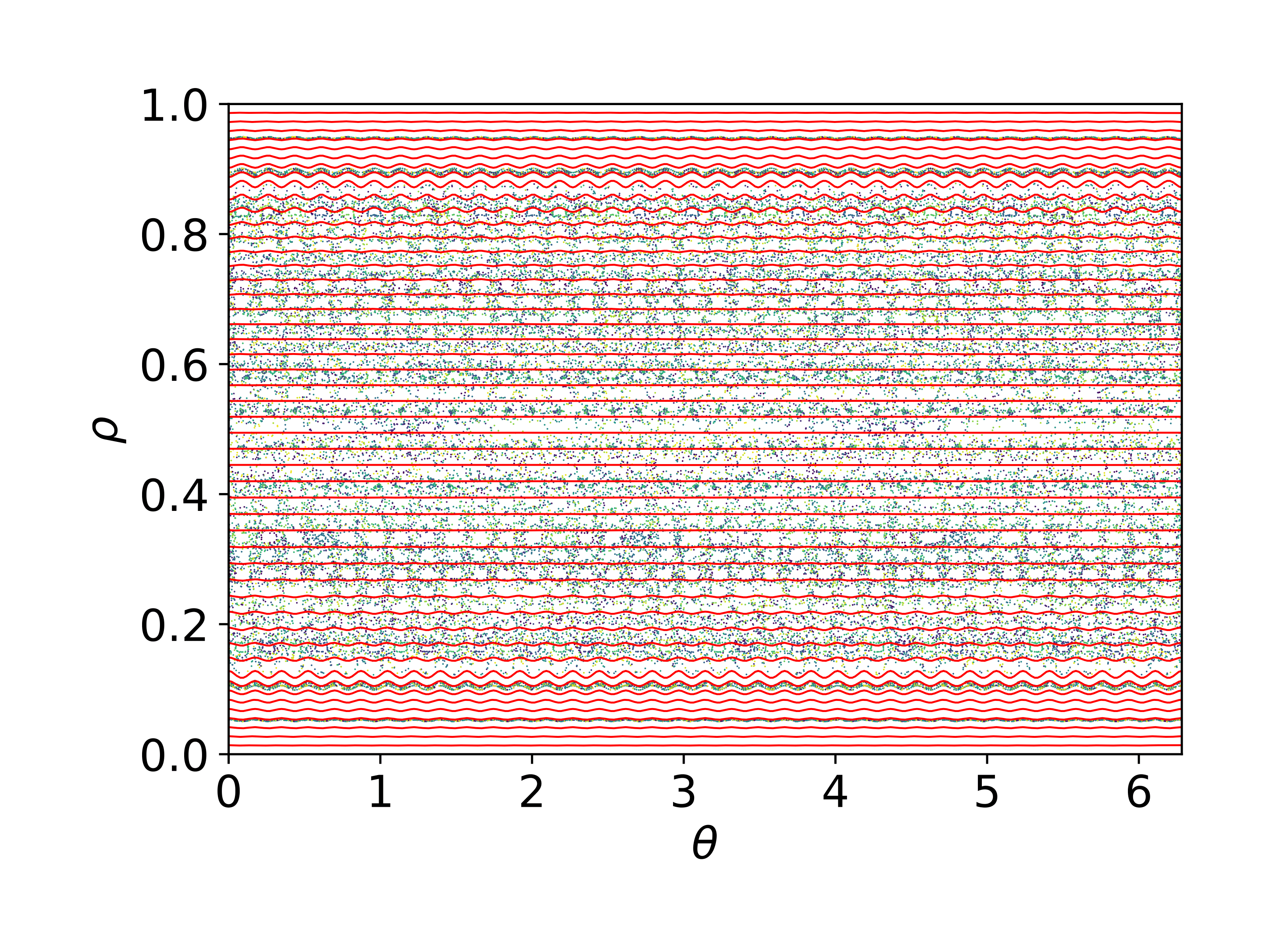}
    \includegraphics[trim=1.2cm 0.8cm 0.8cm 1cm,clip,width=0.48\textwidth]{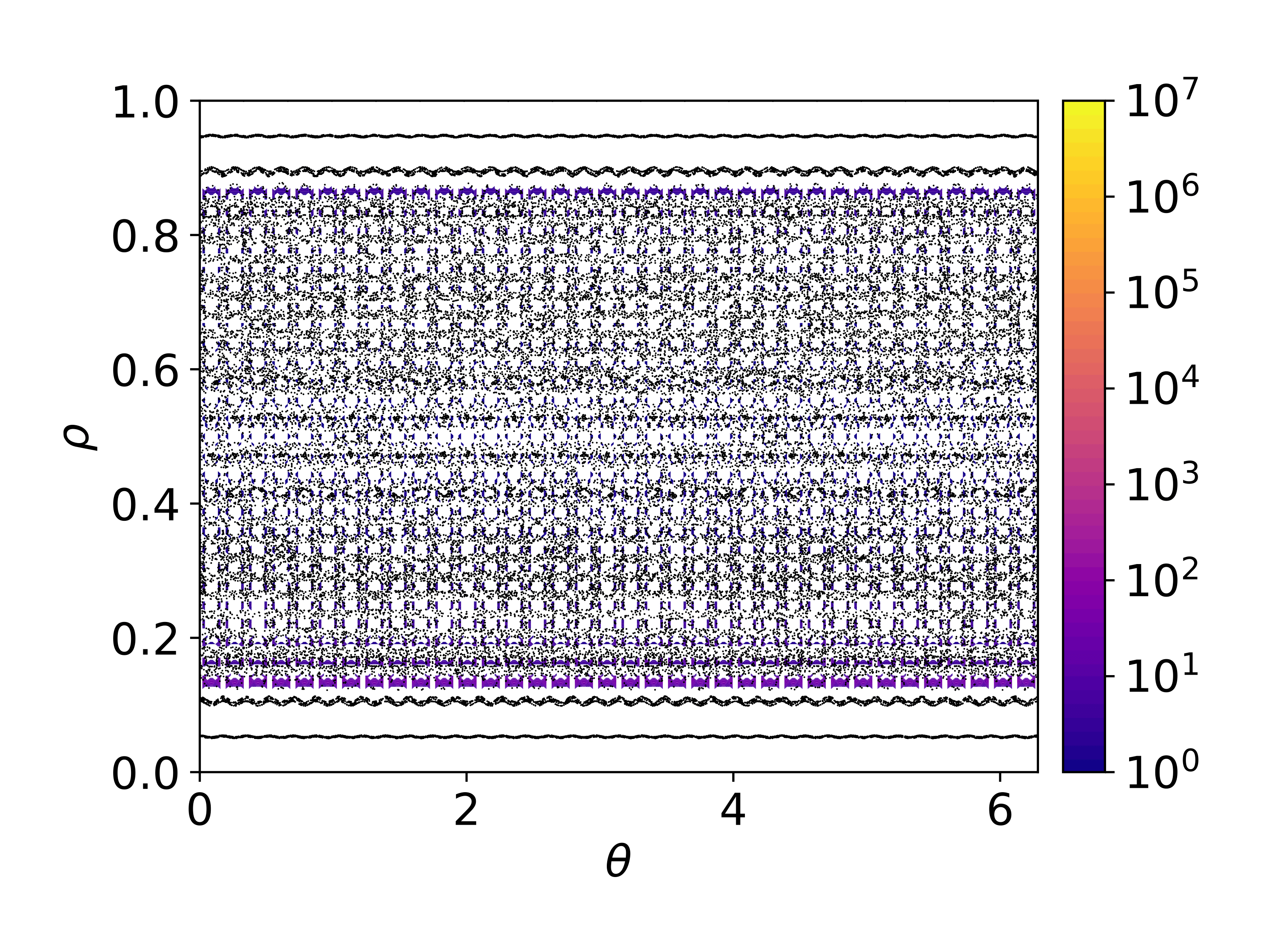}
    \caption{$m = 36$}
    \end{subfigure}
    \caption{The ADE is solved for the model field \eqref{eq:model_field} with critically overlapping resonances of $m = 4$, $m = 12$, and $m=36$ with $\kappa_{\perp} = 10^{-6}$. On the left are the isotherms (red), and on the right is the ratio of the parallel \eqref{eq:parallel_diffusion} to perpendicular \eqref{eq:perpendicular_diffusion} diffusion. The colorscale is {\color{black}set to white} where the parallel diffusion is smaller than the perpendicular diffusion.}
    \label{fig:critical_diffusion_1em6}
\end{figure}

As can be seen from Figure \ref{fig:critical_diffusion_1em6}, each of the fields are visually chaotic with secondary island chains as well as stochastic regions. We furthermore note that the magnitude of the parallel transport, in addition to the volume over which it dominates, is considerably larger in the $m = 4$ field. For the $m = 36$ perturbation, we note that the temperature gradient is not markedly reduced in the chaotic layer when $\kappa_{\perp} = 10^{-6}$, although all of the flux surfaces are destroyed. Thus while all of the fields would be classified as very far from integrability by traditional measures such as the Chirikov overlap criterion or converse KAM theory, the impact of the magnetic field structure on the overall transport at {\color{black}non-zero} $\kappa_{\perp}$ is remarkably distinct. 

\section{Conclusions}
\label{sec:conclusions}

We have proposed a metric, the {\color{black}\textit{effective volume of parallel diffusion}}, which quantifies the fractional volume over which the parallel transport is dominant over the perpendicular transport. This enables one to determine if a magnetic field is sufficiently integrable with regards to its impact on the transport of heat. We validate this metric for two model fields: a single island chain and strongly chaotic layer. We then consider a critically chaotic magnetic field generated by resonances of different mode numbers. We find that even if no KAM surfaces exist, the relative parallel transport can vary widely between these critically chaotic fields, especially at larger values of $\kappa_{\perp}$. This highlights the importance of quantifying non-integrability not only through the stability of a given orbit or the non-existence of a given KAM surface, but through the resulting impact on the transport. We anticipate many applications of this analysis, such as for the optimization of stellarator magnetic fields and analysis of transport in the stochastic edge. 

As discussed in \S \ref{sec:effective_volume}, our metric appears to be related to the converse KAM approach, which enables a quantification of the volume of phase space that does not contain KAM surfaces of a given foliation. We expect that the {\color{black}effective volume of parallel diffusion might} agree with such a calculation in the limit of small perpendicular diffusion. We reserve such a comparison for future publication.


\section*{Acknowledgements}

The authors would like to acknowledge fruitful discussions with D. Pfefferle, P. Constantin, R. Mackay, T. Drivas, D. Ginsberg, and A. Bhattacharjee. 

\section*{Funding}

This work was supported in part by a grant from the Simons Foundation/SFARI (560651) and by DOE Contract No. DEAC02–76CH03073.

\section*{Declaration of interests} 
The authors report no conflict of interest.

\section*{Author ORCID} 
E. J. Paul, https://orcid.org/0000-0002-9355-5595; S. R. Hudson, https://orcid.org/0000-0003-1530-2733; P. Helander, https://orcid.org/0000-0002-0460-590X.

\appendix

\section{Maximum principle for ADE}
\label{app:maximum_principle}


We can now prove a maximum principle for the anisotropic diffusion equation \citep{Fraenkel2000}. Let $T$ satisfy the ADE. We define a function $w = T + \epsilon e^{K \textbf{x} \cdot \textbf{c}}$ for constant scalars $K>0$, $\epsilon>0$, and constant vector $\textbf{c}$. Applying the ADE operator to $w$, we obtain,
\begin{align}
    \nabla \cdot \left(\bm{\kappa} \cdot \nabla w \right) = \epsilon \left(\textbf{c} \cdot \bm{\kappa} \cdot \textbf{c} K^2 + \left(\nabla \cdot \bm{\kappa}\right) \cdot \textbf{c}  K \right) e^{K\textbf{x} \cdot \textbf{c}}.
\end{align}
From the positive definiteness of $\bm{\kappa}$, we note that $\textbf{c} \cdot \bm{\kappa} \cdot \textbf{c} \ge \lambda_0 |\textbf{c}|^2$ where $\lambda_0>0$ is the smallest eigenvalue of $\bm{\kappa}$. This yields the inequality,
\begin{align}
\nabla \cdot \left(\bm{\kappa} \cdot \nabla w \right)
&\ge \epsilon \left( \lambda_0 |\textbf{c}|^2K^2 -  \max_{\Omega} \left|\left(\nabla \cdot \bm{\kappa}\right)\cdot \textbf{c} \right| K \right) e^{K\textbf{x} \cdot \textbf{c}}.
\end{align}
We choose $K > \max_{\Omega} \left|\left(\nabla \cdot \bm{\kappa}\right)\cdot \textbf{c} \right|/(\lambda_0 |\textbf{c}|^2)$ so that $\nabla \cdot \left(\bm{\kappa} \cdot \nabla w \right)>0$.

Suppose now that the maximum value of $w$ is obtained within the domain, $\Omega$, at a point $\textbf{x}_0$. This implies that $\nabla w\rvert_{\textbf{x}_0} = 0$ and the Hessian matrix $\nabla \nabla w \rvert_{\textbf{x}_0}$ is negative semi-definite. This implies that,
\begin{align}
    \nabla \cdot \left(\bm{\kappa} \cdot \nabla w \right)\rvert_{\textbf{x}_0} = \sum_{i,j} \kappa_{ij}(\textbf{x}_0) \partial_i \partial_j w(\textbf{x}_0) = \mathrm{tr}\left(\bm{\kappa} \nabla \nabla w \right)_{\textbf{x}_0} \le 0. 
\end{align}
The final inequality follows from expressing $\bm{\kappa}$ in terms of the eigenvalues, $\mu_i$, and eigenvectors, $\overrightarrow{\textbf{x}}_i$, of $\nabla \nabla w$ with $\mu_i \le 0$,
\begin{align}
    \bm{\kappa} = \sum_i\omega_i \overrightarrow{\textbf{x}}_i \overrightarrow{\textbf{x}}_i^T.
\end{align}
From the positive-definiteness of $\bm{\kappa}$, we find that $\omega_i>0$. We can then note that,
\begin{align}
  \text{tr} \left(\bm{\kappa} \nabla \nabla w \right)_{\textbf{x}_0} = \text{tr}\left( \sum_i \omega_i \mu_i \overrightarrow{\textbf{x}}_i \overrightarrow{\textbf{x}}_i^T \right) \le 0.
\end{align}
As this is in contradiction with the choice we have made for $K$, we conclude that the maximum of $w$ must occur on $\partial \Omega$ and,
\begin{align}
    T < w \le \max_{\overline{\Omega}}(w) = \max_{\partial \Omega} w \le \max_{\partial \Omega} T + \epsilon \max_{\partial \Omega} e^{K \textbf{x} \cdot \textbf{c}}.
\end{align}
This implies that $\max_{\overline{\Omega}}T = \max_{\partial \Omega}T$. Otherwise, we would have $\max_{\Omega}T = \max_{\partial \Omega}T + \delta$ for some $\delta > 0$. As we are free to choose $\epsilon < \delta/\max_{\partial \Omega} e^{K \textbf{x} \cdot \textbf{c}}$, this would lead to a contradiction. A similar argument can be constructed to show that the minimum value of $T$ must occur on the boundary.

\section{Topology of isotherms}
\label{sec:topology}

{\color{black}
In this Appendix, we discuss the possible topology of the isotherms given properties of the ADE. 

To begin, we consider the region near the boundaries.} We first note that the normal derivative of the temperature must be sign-definite on the boundaries unless $T$ is constant in $\Omega$. In other words, $\hat{\textbf{n}} \cdot \nabla T > 0$ on $S_{-}$ and $S_{+}$ assuming $T_{+} > T_{-}$ and $\hat{\textbf{n}}$ is oriented in the direction of increasing $\rho$. This result is known as the Hopf lemma \citep{Fraenkel2000}. Intuitively, the flux through the boundary cannot change sign since $S_{-}$ and $S_{+}$ are isotherms. As a result, in a small neighborhood of the boundary $\partial T/\partial \rho \ne 0$, and isotherms close with the same topology as the boundary. We conclude that at least within a small neighborhood of $S_{-}$ and $S_{+}$, isotherms must be deformable to the boundaries. 

We now argue that the temperature gradient cannot vanish within a subvolume of $\Omega$ if the specified temperature differential $T_{+}-T_{-}$ is non-zero. To see this, suppose that the temperature is constant within a subvolume, $\mathcal{V}$. We remark that under the assumption that the coefficients appearing in the elliptic problem \eqref{eq:elliptic_problem}, $\bm{\kappa}$ and $\nabla \cdot \bm{\kappa}$, are analytic, $T$ must also be analytic due to the ellipticity of the ADE operator. If $T$ is analytic and is constant within $\mathcal{V}$, this implies that $T$ is constant in $\Omega$. Since such a solution is inconsistent with the boundary conditions, we conclude that the temperature gradient can only vanish on isolated lines, surfaces, or points. These must be saddle points given the maximum principle.

{\color{black}
We now consider the possible linking of isothermal surfaces (Figure \ref{fig:isotherm_topologies}). Suppose that there exists an isotherm, $\mathcal{S}$, with the linking of a sphere (i.e., does not close toroidally or poloidally) or the linking of an ``island'' (i.e., closes toroidally or poloidally but not both). In that case, we can integrate the ADE over a volume $\mathcal{V}$ bounded by one of these isotherms, $\mathcal{S} = \partial \mathcal{V}$,
\begin{align}
    \int_{\mathcal{V}} d^3 x \, \nabla \cdot \left(\bm{\kappa} \cdot \nabla T\right) = \int_{\partial \mathcal{V}} d^2 x \, \hat{\textbf{n}} \cdot \bm{\kappa} \cdot \nabla T = 0.
    \label{eq:integration_ade}
\end{align}
We note that $\nabla T \propto \hat{\textbf{n}}$ and $\hat{\textbf{n}} \cdot \nabla T$ is sign-definite on $\partial \mathcal{V}$. Because $\bm{\kappa}$ is positive-definite, $\hat{\textbf{n}} \cdot \bm{\kappa} \cdot \nabla T$ will also be sign-definite on $\partial \mathcal{V}$. This leads to a contradiction unless $\nabla T = 0$. Given that the temperature gradient cannot vanish within a finite volume, we conclude that the isotherms must link as the boundaries such that they do not enclose a net volume.

\begin{figure}
    \centering
    \begin{subfigure}[b]{0.48\textwidth}
    \includegraphics[width=1.0\textwidth]{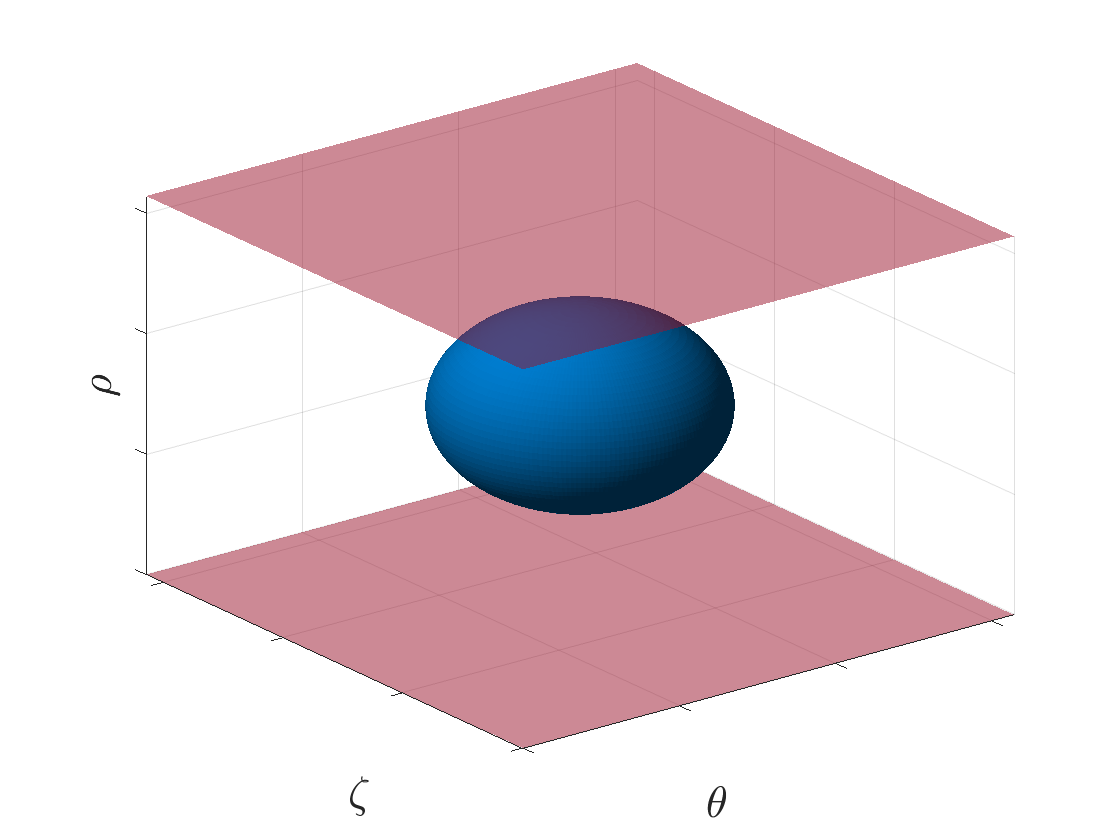}
    \caption{}
    \end{subfigure}
    \begin{subfigure}[b]{0.48\textwidth}
    \includegraphics[width=1.0\textwidth]{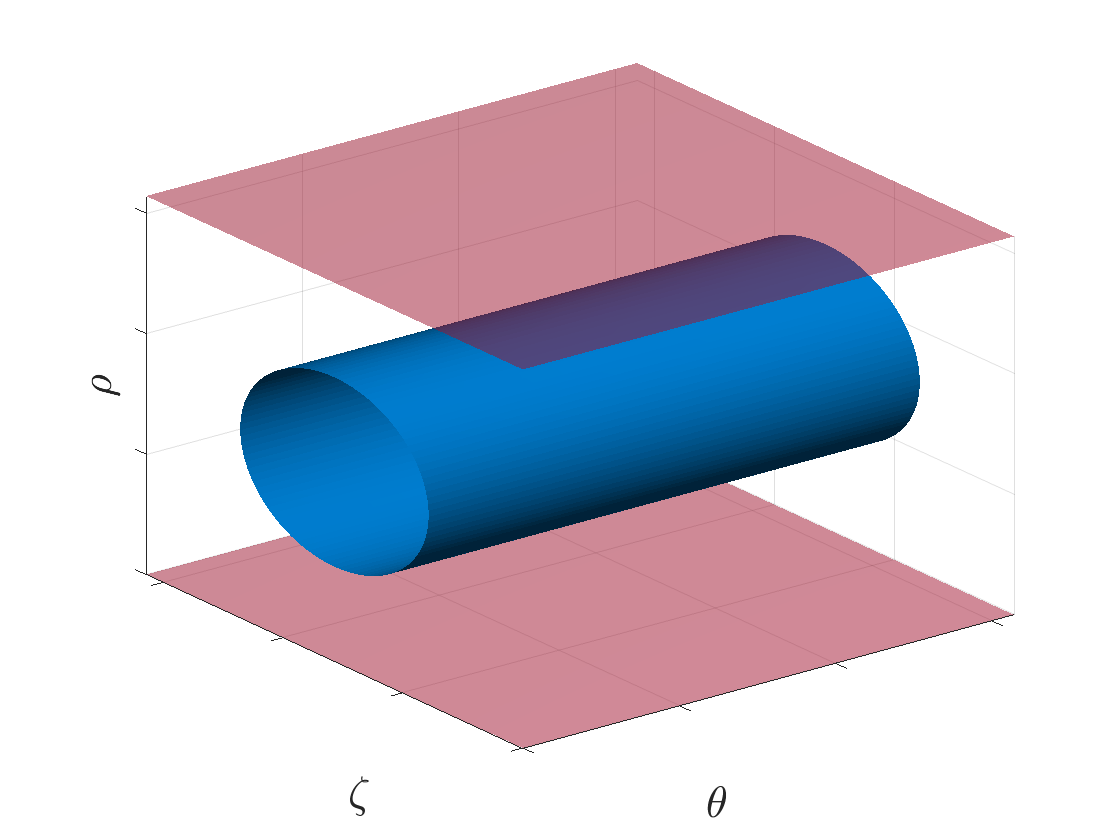}
    \caption{}
    \end{subfigure}
    \begin{subfigure}[b]{0.48\textwidth}
    \includegraphics[width=1.0\textwidth]{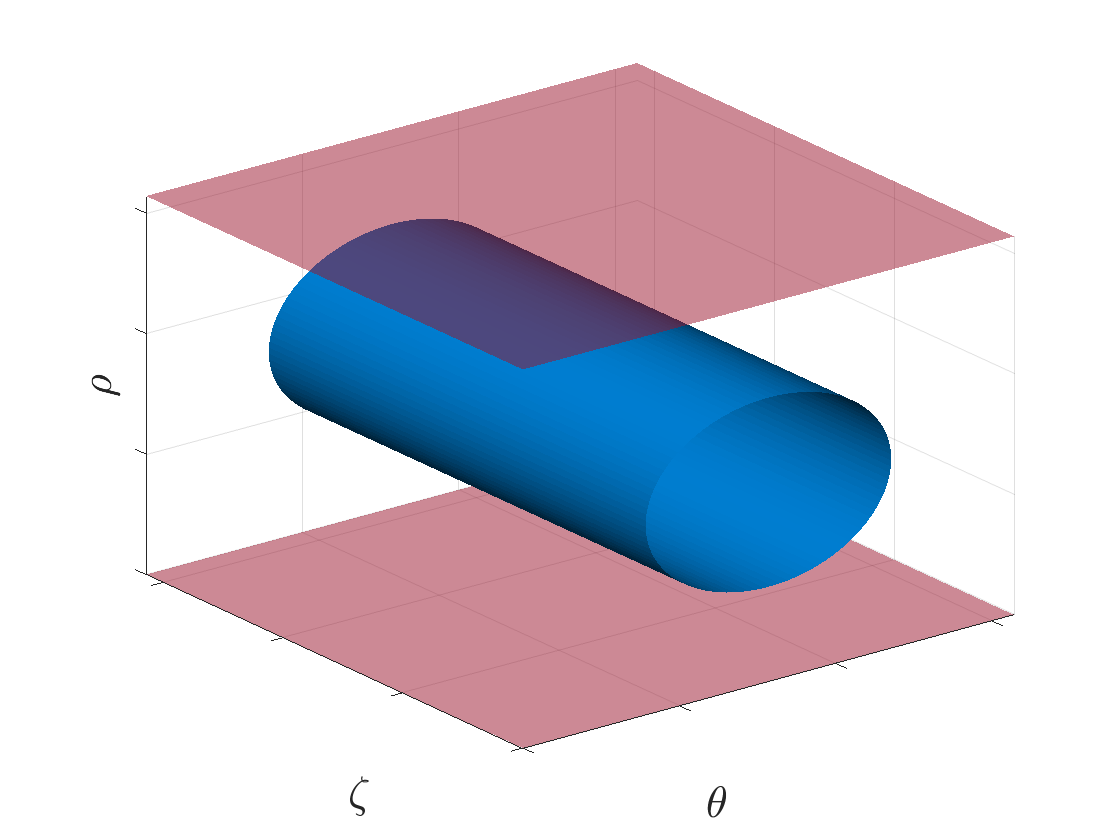}
    \caption{}
    \end{subfigure}
    \begin{subfigure}[b]{0.48\textwidth}
    \includegraphics[width=1.0\textwidth]{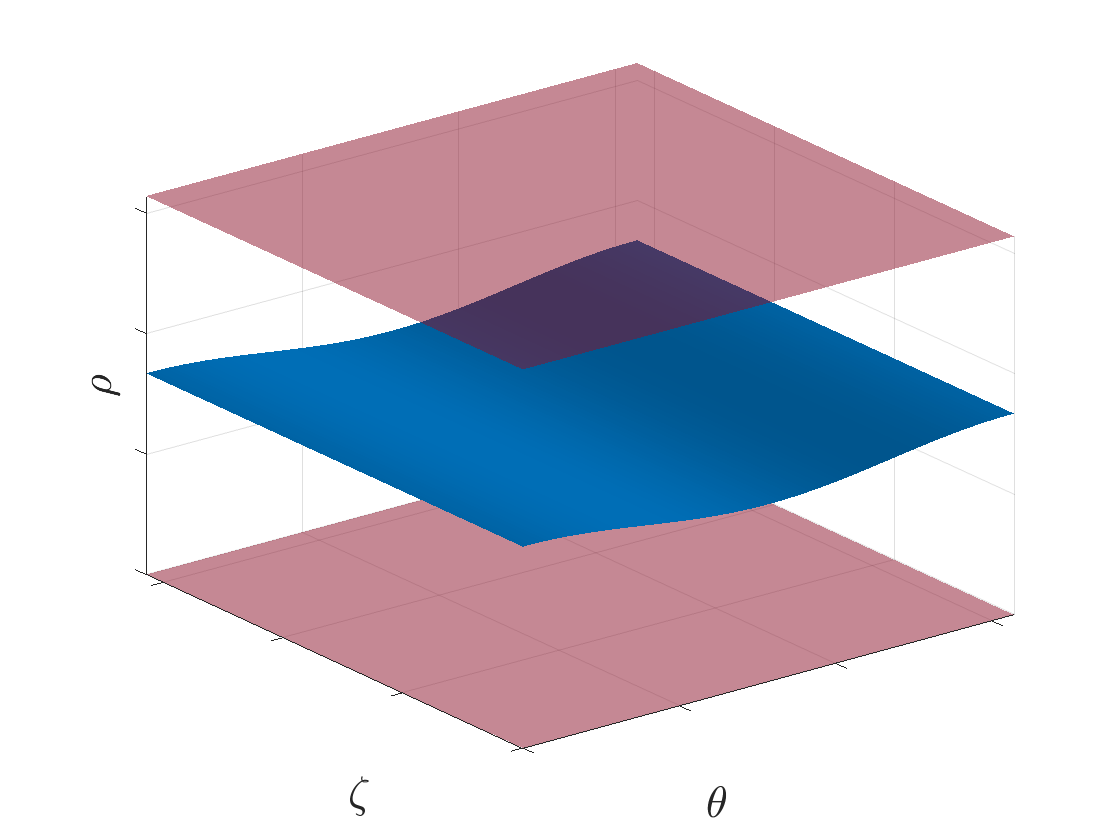}
    \caption{}
    \end{subfigure}
    \begin{subfigure}[b]{0.48\textwidth}
    \includegraphics[width=1.0\textwidth]{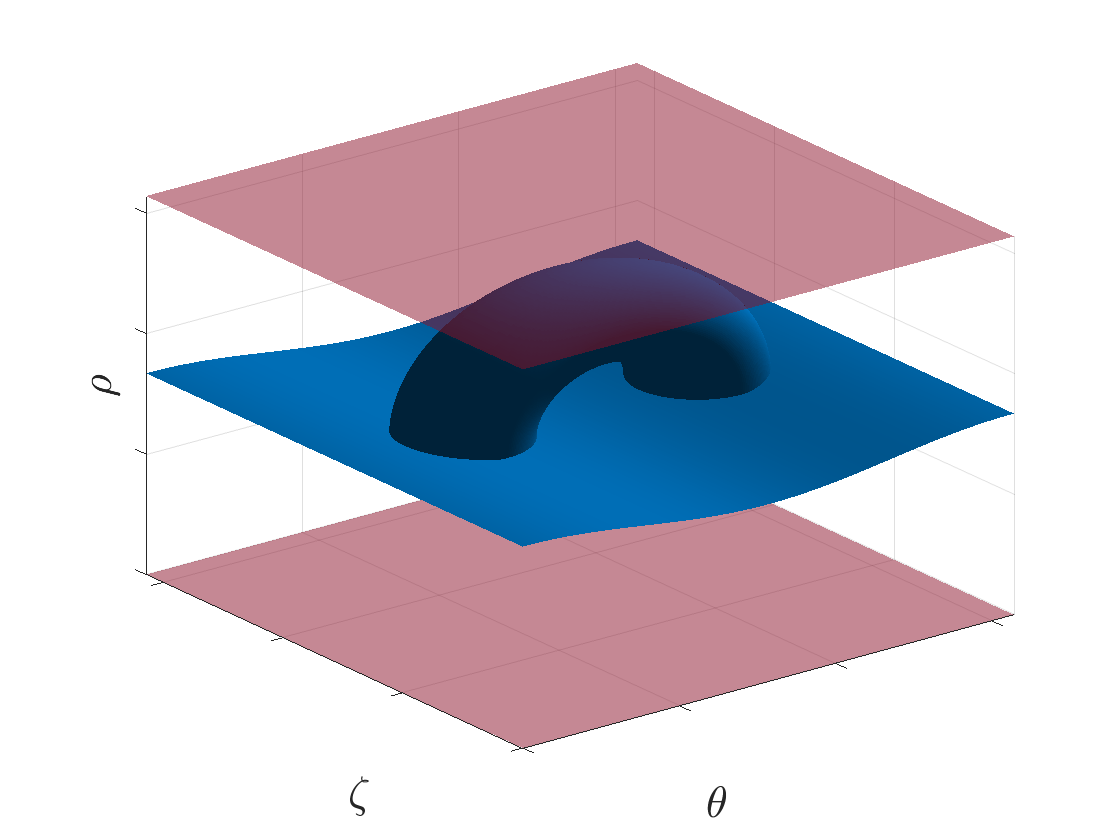}
    \caption{}
    \label{fig:handle}
    \end{subfigure}
    \caption{Possible linking of closed isotherms (blue) which are periodic and do not intersect the boundaries (red): (a) ``spherical'' (not closed toroidally or poloidally), (b) poloidally closed ``island'' structure, (c) toroidally closed ``island'' structure, (d) toroidally and poloidally closed (deformable to boundary), and (e) a toroidally and poloidally closed surface with the addition of a handle.}
    \label{fig:isotherm_topologies}
\end{figure}

Now that we have argued that the isotherms must link toroidally and poloidally as the boundaries, we now discuss their topology. We consider an isotherm which links as the boundaries but with the addition of a small handle (Figure \ref{fig:handle}). Such an isotherm is consistent with the integral condition \eqref{eq:integration_ade} as it does not enclose a volume. We conclude that the isotherms must link toroidally and poloidally, with the possible addition of a handle. However, isotherms cannot enclose a net volume, as an ``island'' or ``sphere'' structure.}

\section{Single island chain analysis}
\label{app:island_analysis}

{\color{black}
In this Appendix we analyze the ADE for the case of a single small island chain, as discussed in Section \ref{sec:single_island}. We perform a perturbation analysis in the small parameter $\epsilon = |\chi_1|/|\chi_0| \ll 1$ where $\chi_0 = \iota' \psi^2/2$ and $\chi_1 = \chi - \chi_0$. In the limit $\chi = \chi_0$, the solution of the ADE is given by,
\begin{align}
    T_0(\rho) = T_- + \frac{\rho(T_+-T_-)}{\overline{\rho}}.
\end{align}
We will call this the unperturbed temperature. In computing the first order correction, the perturbed temperature, we apply the ansatz $T_1(\rho,\theta,\zeta) = \hat{T}(\rho) \cos(m \theta - n \zeta)$. The first order perturbation to the ADE reads,
\begin{align}
    \kappa_{\perp}\nabla^2 T_1 + (1 - \kappa_{\perp})\hat{\textbf{b}}_0 \cdot \nabla\left( \hat{\textbf{b}}_0 \cdot \nabla T_1 + \hat{\textbf{b}}_1 \cdot \nabla T_0\right) = 0,
\end{align}
where $\hat{\textbf{b}}_0$ and $\hat{\textbf{b}}_1$ are the unit vectors in the direction of the magnetic field at $\mathcal{O}(\epsilon^0)$ and $\mathcal{O}(\epsilon^1)$. Applying the ansatz for $T_1$ results in the ODE,
\begin{multline}
     \kappa_{\perp} \left( \hat{T}''(\rho) - \hat{T}(\rho) \left(\frac{m^2}{\overline{\rho}^2} + \frac{n^2}{ L_{\zeta}^2} \right) \right)  \\
     - (1 - \kappa_{\perp}) \frac{m^2 \iota'^2 \overline{\psi}^2}{L_{\zeta}^2 \overline{\rho}^2\left(1 + \frac{n^2\overline{\rho}^2}{m^2  L_{\zeta}^2}\right)} \left(\hat{T}(\rho) \left(\rho - \rho_{\text{r}} \right)^2
    -  \frac{\mathcal{W}_{m,n}^2 (\rho - \rho_{\text{r}})}{4\overline{\rho}} (T_+-T_-)\right) = 0,
\end{multline}
where $\mathcal{W}_{m,n}$ is the island half-width \eqref{eq:island_half_width}. In obtaining the above expression, we focus on a region near the $n/m$ rational surface such that the unperturbed field strength can be approximated as $|\textbf{B}_0| = \left(\overline{\psi}/\overline{\rho}^2\right) \left(1 + \left(\iota' \overline{\psi} \rho/L_{\zeta}\right)^2\right)^{1/2} \approx\left(\overline{\psi}/\overline{\rho}^2\right) \left(1 + \left(n \overline{\rho}/(mL_{\zeta})\right)^2\right)^{1/2} $.
}
{\color{black}
Due to the smallness of $\kappa_{\perp}$, outside of a small boundary layer of width $\mathcal{W}_{\text{c}}$, the perpendicular diffusion becomes unimportant. We can express the first order perturbation to the ADE in terms of the layer variable $X = (\rho - \rho_{\text{r}})/\mathcal{W}_{\text{c}}$, where $\rho_{\text{r}} = \overline{\rho}n/(m \iota'\overline{\psi})$ is the location of the resonant surface,
\begin{multline}
     \kappa_{\perp} \left( \frac{\hat{T}''(X)}{\mathcal{W}_{\text{c}}^2} - \hat{T}(X) \left(\frac{m^2}{\overline{\rho}^2} + \frac{n^2}{ L_{\zeta}^2} \right) \right)  \\
     - (1 - \kappa_{\perp}) \frac{m^2 \iota'^2 \overline{\psi}^2}{L_{\zeta}^2 \overline{\rho}^2\left(1 + \frac{n^2\overline{\rho}^2}{m^2  L_{\zeta}^2}\right)} \left(\hat{T}(X) \mathcal{W}_{\text{c}}^2 X^2 
    -  \frac{\mathcal{W}_{m,n}^2 \mathcal{W}_{\text{c}} X}{4\overline{\rho}} (T_+-T_-)\right) = 0.
    \label{eq:boundary_layer}
\end{multline}
We will analyze the solution both inside and outside of the layer, similar to the analysis in \cite{Fitzpatrick1995}. 

\subsection{Inner solution}

Inside the boundary layer, the second term in \eqref{eq:boundary_layer} is unimportant as $\mathcal{W}_{\text{c}}/\overline{\rho} \ll 1$ by assumption. Thus the first and third terms must balance each, allowing us to obtain the scaling for the boundary layer width as \eqref{eq:boundary_width}. Balance between the first and last terms yields a scaling for the temperature perturbation as,
\begin{align}
    \hat{T}(X) \sim \frac{\mathcal{W}_{m,n}^2}{\overline{\rho} \mathcal{W}_{\text{c}}} (T_+ - T_-). 
\end{align}
This matches the scaling in (32) of \cite{Fitzpatrick1995}. This implies that $|\nabla_{\perp} T_1|  \sim  \left(\mathcal{W}_{m,n}/\mathcal{W}_{\text{c}}\right)^2 |\nabla_{\perp} T_0| $. 
When $\mathcal{W}_{m,n} \ll \mathcal{W}_{\text{c}}$, $|\nabla_{\perp} T_0| \gg |\nabla_{\perp} T_1|$, and the ratio of parallel to perpendicular diffusion scales as,
\begin{align}
    \frac{|\nabla_{\|} T|^2}{\kappa_{\perp}|\nabla_{\perp} T|^2} \sim \frac{\mathcal{W}_{m,n}^4}{\mathcal{W}_{\text{c}}^4} \ll 1,
\end{align}
since $|\nabla_{\|} T_1| \sim \mathcal{W}_\text{c}^{-1} \kappa_{\perp}^{1/2} T_1$.
Otherwise, if $\mathcal{W}_{m,n}/\mathcal{W}_{\text{c}} \gtrsim 1$, the ratio can be $\sim 1$.

\subsection{Outer solution}

Outside of the boundary layer, the perpendicular diffusion can be ignored to a good approximation, and the ADE reduces to,
\begin{align}
    \nabla \cdot \left(\hat{\textbf{b}} \hat{\textbf{b}} \cdot \nabla T \right) = 0.
    \label{eq:outer_pde}
\end{align}
Outside of the island separatrix, we apply a perturbation analysis to obtain,
\begin{align}
    \hat{\textbf{b}}_0 \cdot \nabla \left( \hat{\textbf{b}}_0 \cdot \nabla T_1 + \hat{\textbf{b}}_1 \cdot \nabla T_0 \right) = 0.
    \label{eq:outside_ode}
\end{align}
This yields the solution,
\begin{align}
    \hat{T}^{(0)}(\rho) = \mathcal{W}_{m,n}^2\frac{T_+-T_-}{4\overline{\rho} (\rho - \rho_{\text{r}})},
    \label{eq:outer_solution}
\end{align}
where the superscript $(0)$ indicates the solution which neglects perpendicular diffusion. This matches the scaling of (38) in \cite{Fitzpatrick1995}. Given \eqref{eq:outside_ode}, the parallel temperature gradient is negligible in this region if perpendicular diffusion is ignored. Now retaining the perpendicular diffusion terms from \eqref{eq:boundary_layer}, the correction to $\hat{T}(\rho)$ scales as $\hat{T}^{(1)}(\rho)\sim \hat{T}(\rho) \mathcal{W}_{\text{c}}^4/\left(\overline{\rho}^2 (\rho - \rho_{\text{r}})^2\right)$, and the parallel diffusion will scale as $|\nabla_{\|} T| \sim \kappa_{\perp}^{1/2} \mathcal{W}_{\text{c}}^{-2} (\rho - \rho_{\text{r}}) \hat{T}^{(1)}(\rho)$. Given that $|\nabla_{\perp} T_0| \gg |\nabla_{\perp} T_1|$ in this region, the ratio of the parallel to perpendicular diffusion will have the scaling,
\begin{align}
    \frac{|\nabla_{\|} T|^2}{\kappa_{\perp}|\nabla_{\perp} T|^2} \sim \frac{\mathcal{W}_{m,n}^4 \mathcal{W}_{\text{c}}^4 }{\overline{\rho}^4(\rho - \rho_{\text{r}})^4} \ll 1.
\end{align}

Within the island separatrix, \eqref{eq:outer_pde} implies that the temperature must be a function of the island flux surfaces. This is not possible unless the temperature is flattened across the island chain given the maximum principle. 
At non-zero $\kappa_{\perp}$, a finite temperature gradient is allowed. 
Given that the temperature is flattened within the separatrices if perpendicular diffusion is not included, both the parallel and perpendicular temperature gradients will be driven by $\hat{T}^{(1)}(\rho)$ with scaling given by $|\nabla_{\|} T| \sim \kappa_{\perp}^{1/2} \mathcal{W}_{\text{c}}^{-2} (\rho - \rho_{\text{r}})\hat{T}^{(1)}(\rho)$ and $|\nabla_{\perp} T| \sim \hat{T}^{(1)}(\rho)/(\rho - \rho_{\text{r}})$. This yields the scaling for the ratio of the parallel to perpendicular diffusion as,
\begin{align}
    \frac{|\nabla_{\|} T|^2}{\kappa_{\perp}|\nabla_{\perp} T|^2} \sim \frac{(\rho - \rho_{\text{r}})^4}{\mathcal{W}_{\text{c}}^4} \gg 1.
\end{align}

We conclude that for the outer solution, the ratio $|\nabla_{\|}T|^2/\left(\kappa_{\perp}|\nabla_{\perp}T|^2\right)$ can only be $\gtrsim 1$ if $|\rho-\rho_{\text{r}}|< \mathcal{W}_{m,n}$ and  $\mathcal{W}_{\text{c}} \ll \mathcal{W}_{m,n}$.

}

\section{Strong chaos analysis}
\label{app:chaos_analysis}

{\color{black}
In this Appendix we analyze the ADE for the case of strong chaos, as discussed in Section \ref{sec:strong_chaos}.
} If we assume that $\epsilon_{m,n}$ is sufficiently small, we can linearize the trajectories {\color{black}with respect to the small parameter $\epsilon = |\chi_1|/|\chi_0| \ll 1$ where $\chi_0 = \iota' \psi^2/2$ and $\chi_1 = \chi - \chi_0$},
\begin{subequations}
\begin{align}
    \psi &= \psi_0 + \psi_1(\zeta) + \mathcal{O}(\epsilon^2)\\
    \theta &= \theta_0 + \iota(\psi_0) \zeta + \theta_1(\zeta) + \mathcal{O}(\epsilon^2),
\end{align}
\end{subequations}
where {\color{black}$\iota(\psi_0) = \iota' \psi_0$} is the rotational transform associated with the unperturbed Hamiltonian \eqref{eq:model_field}. The linearized trajectories satisfy,

\begin{subequations}
\begin{align}
    \der{\psi_1}{\zeta} &= \sum_{m,n} m \widetilde{\epsilon}_{m,n}(\psi_0)  \sin(m (\theta_0 + \iota(\psi_0) \zeta ) - n \zeta) \\
    \der{\theta_1}{\zeta} &= \iota'\psi_1 + \sum_{m,n}\widetilde{\epsilon}_{m,n}'(\psi_0) \cos(m (\theta_0 + \iota(\psi_0) \zeta) - n \zeta),
\end{align}
\end{subequations}
and are computed to be,
{\color{black}
\begin{subequations}
\begin{align}
    \psi_1(\zeta) &= -\sum_{m,n} m\widetilde{\epsilon}_{m,n}(\psi_0) \frac{\cos(m (\theta_0 + \iota(\psi_0) \zeta) - n \zeta)}{m \iota(\psi_0) - n} \\
    \theta_1(\zeta) &= \sum_{m,n} \left[ \frac{-m \iota' \widetilde{\epsilon}_{m,n}(\psi_0)}{(m \iota(\psi_0) - n)^2} + \frac{\widetilde{\epsilon}_{m,n}'(\psi_0)}{m \iota(\psi_0) - n} \right]\sin(m (\theta_0 + \iota(\psi_0) \zeta) - n \zeta).
\end{align}
\end{subequations}
}
Thus we can see that the resonant terms in $\theta_1(\zeta)$ within a width $\mathcal{W}_{m,n}/2$ \eqref{eq:island_half_width} of a given rational surface will be $\mathcal{O}(1/m)$ and will  thus dominate the evolution. We can then approximate the trajectories as,
{\color{black}
\begin{subequations}
\begin{align}
    \der{\psi_1}{\zeta} &\approx \sum_{m,n}\Theta(\mathcal{W}_{m,n}/2 -|\rho-\rho_{\text{r}}|) m \widetilde{\epsilon}_{m,n}(\psi_0) \sin(m \theta_0 ) \\
    \der{\theta_1}{\zeta} &\approx \iota'\psi_1 + \sum_{m,n}\Theta(\mathcal{W}_{m,n}/2 -|\rho-\rho_{\text{r}}|) \widetilde{\epsilon}_{m,n}'(\psi_0) \cos(m \theta_0),
\end{align}
\end{subequations}
}
where $\Theta$ is the Heaviside step function. In one toroidal period, the flux changes by an amount given by {\color{black} $\Delta \psi \approx 2\pi d \psi_1/ d \zeta $}
and the poloidal angle changes by an amount given by {\color{black}$\Delta \theta \approx 2\pi d \theta_1/ d \zeta $}.
We note that the uncertainty in the poloidal angle will be $\mathcal{O}(1)$ if the {\color{black}resonances} are strongly overlapping. In this way, with every toroidal period, the poloidal angle will be kicked by an amplitude that depends on the surrounding resonances. We assume that $\theta_0$ can, therefore, be treated as a random variable, and we approximate the motion in such a field as a random walk,
\begin{align}
    \partder{T}{\zeta} \sim D_{\|} \partder{^2T}{\psi^2},
\end{align}
with a diffusion coefficient given by,
{\color{black}
\begin{align}
  D_{\|} \sim \frac{\langle (\Delta \psi)^2 \rangle}{2 \Delta \zeta},
\end{align}
}
where $\langle ... \rangle$ indicates an average over the possible initial conditions, $\theta_0$. {\color{black}
The full expression for $D_{\|}$ is given in  \eqref{eq:diffusion_coeff}.
}

\bibliographystyle{jpp}
\bibliography{jpp-instructions}

\end{document}